\documentclass[aps,prx,twocolumn]{revtex4-1}
\usepackage{graphicx}
\usepackage[bottom]{footmisc}

\usepackage{amsmath}
\usepackage{amssymb}
\usepackage{braket}
\usepackage{hyperref}
\usepackage{subfigure}
\usepackage[usenames,dvipsnames]{color}
\usepackage{graphicx}
\usepackage{subfigure}
\usepackage{cleveref}

\newcommand{\ct}{\cite}

\newcommand{\bi}{\bibitem}
\newcommand{\be}{\begin{equation}}
\newcommand{\ee}{\end{equation}}
\newcommand{\ba}{\begin{eqnarray}}
\newcommand{\ea}{\end{eqnarray}}

\begin{document}
\begin{@twocolumnfalse}

\title{Exploring the possibilities of dynamical quantum phase transitions  in the presence of a Markovian bath}
\author{Souvik Bandyopadhyay}
\email{souvik@iitk.ac.in}
\affiliation{Department of Physics, Indian Institute of Technology, Kanpur, Kanpur 208016, India}
\author{Sudarshana Laha}
\affiliation{Department of Physics, Indian Institute of Technology, Kanpur, Kanpur 208016, India}
\author{Utso Bhattacharya}
\affiliation{Department of Physics, Indian Institute of Technology, Kanpur, Kanpur 208016, India}
\author{Amit Dutta}
\affiliation{Department of Physics, Indian Institute of Technology, Kanpur, Kanpur 208016, India}

	\begin{abstract}
	We explore the possibility of dynamical quantum phase transitions (DQPTs) occurring during the temporal evolution of  a quenched transverse field Ising chain coupled to a particle loss type of bath (local in Jordan-Wigner fermion space) using
	two versions of the Loschmidt overlap (LO), namely, the fidelity induced LO and the interferometric phase induced LO. The bath, on the one hand, dictates the dissipative evolution following 
	a sudden quench and on the other, plays a role in dissipative mixed state preparation in the later part of the study. During a dissipative evolution following a sudden quench, no trace of 
	DQPTs are revealed in both the fidelity and the interferometric phase approaches; however, remarkably the interferometric phase approach reveals the possibility of inter-steady state DQPTs in passage from one steady state to the other
	when the system is subjected to a quench after having reached the first steady state. We further probe the occurrences of DQPTs when the system evolves unitarily after being prepared in
	a mixed state of engineered purity by ramping the transverse field in a linear fashion in the presence of the bath. In this case though the fidelity approach fails to indicate any DQPT,  the interferometric approach indeed
	unravels the possibility of occurrence of  DQPTs which persists even up to a considerable loss of purity of the engineered initial  state  as long as a constraint relation involving  the dissipative coupling and ramping time (rate) is satisfied. This constraint relation also marks the boundary between two  {\it dynamically} inequivalent phases; in one  the LO vanishes for the critical momentum mode (and hence DQPTs exist)   while in the other no
	such critical mode   can exist and hence the LO never vanishes.  
		
	\end{abstract}
	\maketitle
\end{@twocolumnfalse}
 
 \section{Introduction}
  
  \label{sec_intro}
  
The study of dynamics of quantum many-body systems driven out of equilibrium is a frontier area of recent research both from the experimental as well as the theoretical viewpoints \ct{greiner02,kinoshita06,gring12,trotzky12,cheneau12,schreiber15,fausti11,rechtsman13,calabrese06,rigol08,bermudez09,
oka09,kitagawa10,lindner11,gambassi11,patel13,caio15,thakurathi13,mukherjee09,das10,Russomanno_PRL12,sharma14,nag14,bukov16,sen16,pal10,nandkishore15}. {(For review articles, we refer to \ct{dziarmaga10,polkovnikov11,dutta15,eisert15,alessio16,jstat}).} 
One of the emerging areas in this regard is understanding so called  dynamical quantum phase transitions (DQPTs) introduced by Heyl $et~al.$ \ct{heyl13}. In a DQPT, non-analyticities
manifest in the subsequent  real-time dynamics of a quantum many-body system generated by the
time-independent final Hamiltonian following  a sudden or a slow ramping of one of the parameters of the Hamiltonian.  An analogy can be drawn between DQPTs and  equilibrium classical phase transitions by analysing the lines of Fisher zeros in the  one-dimensional situation \ct{fisher65}, (see also \ct{lee52,saarloos84})   in the complex time plane. Remarkably, these non-analyticities have been detected experimentally in a string of ions simulating interacting transverse field Ising models \ct{jurcevic16}.  At the same time, the time evolution of a fermionic quantum gas  in  a hexagonal optical lattice, after a rapid quench from a topologically trivial system into a Haldane-like system (by quenching between a static and a Floquet Hamiltonian), has been studied and occurrence of DQPTs  has been ascertained using  momentum- and time-resolved state tomography \ct{flaschner}.\\

Focussing on the one dimensional situation,  let us assume that the system is  prepared in the ground state $|\psi(0)\rangle$ at time $t=0$ when a parameter $\lambda$ of the Hamiltonian in suddenly changed  from an initial value $\lambda_i$ to  a final value  $\lambda_{f}$. DQPTs occur at those instants of time $t$ when the evolved state  $|\psi(t)\rangle= \exp({-iH(\lambda_{f})t)}|\psi(0)\rangle$, where $H(\lambda_f)$ is time-independent final Hamiltonian, becomes orthogonal to the initial state $|\psi(0)\rangle$, i.e., the so-called Loschmidt overlap (LO), ${\cal L}(t)= \langle \psi_0|\psi(t)\rangle $ vanishes. At those instants of time, the so-called  dynamical free energy density (or the rate function of the return probability) defined as ${\cal F}
=-(1/L) \ln |{\cal L}(t)|$ shows cusp-like singularities.\\

Following the initial proposal \ct{heyl13}, there have been a plethora of studies investigating intricacies of DQPTs in several  integrable and non-integrable, one dimensional  (as well as two dimensional)  closed quantum systems occurring subsequent to  a sudden change \ct{karrasch13,kriel14,andraschko14,canovi14,heyl14,vajna14,
vajna15,schmitt15,palami15,heyl15,budich15,huang16,divakaran16,puskarov16,zhang16,heyl16,zunkovic16,sei17,fogarty17,heyl18,bhattacharya1,bhattacharya2,mera17,halimeh17,homri17,dutta17,sedlmayr181,bhattacharjee18,kennes18}. However,
similar non-analyticities do also appear when the initial state is prepared using a slow ramping protocol \ct{pollmann10,sharma15,sharma16}. The possibility of the occurrence
\ct{bandyopadhyay17,heyl_mixed17} or the disappearance \ct{abeling16,sedlmayr182} of DQPTs when a system is initially prepared in a mixed state, rather than a pure state, have also been reported. 
(For reviews on various aspects of DQPTs, we refer to \ct{heyl17,victor17,zvyagin17}.)  On the contrary, the fate of DQPTs for an open quantum system coupled to a bath is relatively less studied. Exploiting  a fidelity based measure of the LO, a recent result \ct{sedlmayr182} suggested that DQPTs following a sudden quench are likely to disappear in the presence of a bath  except for a very fine-tuned situation.

In this work, we investigate DQPTs in a transverse  field Ising chain which is coupled to a Markovian bath local in Jordan-Wigner (JW) space. In the first approach, the initial pure state density matrix is subjected to a sudden quench and the  resulting dissipative post-quench dynamics  in the presence of the bath is analysed. On the contrary, in the second approach  the  {\it initial mixed state} density matrix is prepared
through a linear ramping of the transverse field from an initial to a desired final  value in the presence of dissipation.  The bath is then decoupled and one probes the signature of DQPTs
in the subsequent {\it unitary} evolution generated by the final time-independent Hamiltonian. We note that decoupling of the bath to study the subsequent unitary evolution
of a thermal state turns out to be  relevant in the context of entropy generation in  quenched closed quantum systems \ct{dorner12,sharma_entropy15}.

 At the outset, let us summarise the problem addressed and the relevant results presented in the later sections along with  a discussion on the organisation of the paper. In the first part, we have studied the non-equilibrium dynamics of a one-dimensional  transverse field Ising model (TFIM) coupled to a Markovian, particle-loss type bath which acts locally on the JW fermionic space. We  recall that a similar bath has been used in the context of heat generation (residual energy) in a dissipative linear ramping protocol \ct{fazio17}.
 
 Throughout the paper, we shall make recourse to two  versions of the LO: (i)  the fidelity induced and (ii) the interferometric phase induced, as  introduced in (Sec. \ref{sec_LO}) where we also point 
 to the fact that these measures satisfy all the properties of a metric and both correctly reduce to the pure state limit. In the beginning of (Sec.~\ref{sec_sudden}), 
 we look for the signature of DQPTs during the dissipative dynamics subsequent to a sudden change in the external magnetic field of the TFIM across a quantum critical point (QCP). It is then analytically established that the fidelity induced LO, being of positive semi-definite nature, is unlikely  to capture dissipative DQPTs.  Similarly, considering the interferometric phase induced LO (Sec.~\ref{sec_sudden_interfero}),  we show that the system moves away from dynamical criticality as it evolves towards the dissipative steady state and hence no DQPTs are observed. On the contrary, we establish that remarkably,  after a second quench starting  from the first steady state, as discussed in the (Sec.~\ref{sec_sudden_steady_state}), the system approaches another steady state with possible occurrences of DQPTs in the inter-steady state dynamics. These DQPTs are manifested only in the interferometric phase induced LO which preserves the Bloch sphere structure unlike the fidelity induced LO. The occurrence of this inter-steady state DQPT is unique
 to our study and to the best of our knowledge has not been reported before.  \\
 
 In (Sec.~\ref{sec_unitary}), we digress to the slow ramping situation and  study a unitary evolution of a TFIM following a dissipative state preparation to probe the effect of an {\it initial} state of engineered purity on the unitary DQPTs. To prepare the desired mixed state, the external magnetic field of the transverse field is slowly ramped in the presence of a bath (Sec.~\ref{sec_unitary_preparation}) and the subsequent unitary evolution is probed. In this case  as well,  for the fidelity induced LO, we analytically establish that depending on the time (rate) of the ramping and the time scale of the action of dissipation, the DQPTs are  smoothened out (Sec.~\ref{sec_unitary_fidelity}). However, in the interferometric approach we observe that the DQPTs persist even up to a considerable loss of purity of the {\it initial} state  as long as the dissipative coupling and ramping time (rate) satisfy a constraint relation (Sec.~\ref{sec_unitary_interfero}) in terms of the
 population in different levels; this constraint relation, as we argue, marks the boundary between two {\it dynamically} inequivalent phases . This is in sharp contrast with the entirely unitary DQPTs following a sudden \ct{heyl13} or slow ramping \ct{sharma16}  or even for  sudden quench starting  from a mixed state \ct{bandyopadhyay17,heyl_mixed17}; in all  these cases no such bound exists. Finally, the experimental possibilities and the concluding comments are presented in (Secs.~\ref{sec_expt}) and (\ref{sec_conclusion}), respectively.
 
\section{The relevant metrics and associated Loschmidt overlaps}
\label{sec_LO}

Let  us recall that a metric space can be defined as an ordered pair $(S,d)$, where $S$ is a set and $d$ is a metric on $S$;  one  then defines a  notion of distance function given by
\begin{equation*}
d:S\times S\rightarrow \mathcal{R},
\end{equation*}
(where $\mathcal{R}$ denotes the set of real numbers) such that for any $x,y,z \in S$, the conditions listed below hold true:
\begin{subequations}\label{cond}
	\begin{equation}
		d(x,y)\geq 0 \quad \text{non-negativity};
	\end{equation}
		\begin{equation}
		d(x,y)=1 \implies x=y \quad \text{identity of indiscernibles};
		\end{equation}
			\begin{equation}
			d(x,y)= d(y,x)\quad \text{symmetry};
			\end{equation}
				\begin{equation}
				d(x,z)\leq d(x,y)+d(y,z)\quad\text{triangle inequality}.
				\end{equation}
\end{subequations}
In the subsequent discussions, DQPTs are studied through the zeros of the  LOs constructed out of two different metrics, the density matrix fidelity and the interferometric distance.
The fidelity between two density matrices $\rho$ and $\sigma$  defined as \ct{mixed_fidelity},
\begin{equation}\label{eq:fidelity}
F(\rho,\sigma)=\mathrm{Tr}\left[\sqrt{\sqrt{\sigma}\rho\sqrt{\sigma}}\right],
\end{equation}
necessarily satisfies all the conditions given in Eq.~\eqref{cond}. \ {For a quantum system evolving in time, one can thus obtain a fidelity-based LO}

\be
{\cal L}^F(\rho(0),\rho(t)) = \mathrm{Tr} \left[\sqrt{\sqrt{\rho(t)}\rho(0) \sqrt{\rho(t)}}\right],
\label{eq_LO_fid}
\ee
{where $\rho(0)$ is the initial density matrix and $\rho(t)$ its time evolved counterpart.}

On the other hand, in the interferometric phase approach a pure state version (purification) \ct{uhlmann86,uhlmann89}, $|w\rangle_{\rho} \in \mathcal{H}_{w}$ $(\mathcal{H}_{S}\otimes\mathcal{H}_{A})$ of  a {mixed
state} density matrix $\rho$ is defined as 
\begin{equation}\label{eq:ip}
|w\rangle_{\rho}=\sum_{i}\sqrt{p_{i,\rho}}|\psi_{i}\rangle_{\rho}\otimes |\alpha_{i}\rangle 
\quad \text{where} \quad |\alpha_{i}\rangle \in \mathcal{H}_{A},
\end{equation}
where $\mathcal{H}_{S}$ is the Hilbert space of the system states and $\mathcal{H}_{A}$ is that of the ancillary states and $p_{i,\rho},~ |\psi_{i}\rangle_{\rho}$ are the eigenvalues and eigenvectors of $\rho$ respectively.  The original density matrix can be recovered by partially tracing out the ancillary states as
\begin{equation*}
	\rho= \mathrm{Tr}_{A}\left[_{\rho}|w\rangle\langle w|_{\rho}\right].
\end{equation*}
In a similar spirit,  for a different density matrix $\sigma$, one rewrites  Eq.~\eqref{eq:ip} in the form:

\begin{equation*}
|w\rangle_{\sigma}=\sum_{i}\sqrt{p_{i,\sigma}}|\psi_{i}\rangle_{\sigma}\otimes|\alpha_{i}^{\prime}\rangle.
\end{equation*}
Equipped with the definition of purification, one can proceed to define the interferometric distance \ct{sjoqvist} between density matrices $\rho$ and $\sigma$
 \begin{equation}\label{eq:25}
 d(\rho,\sigma)=|_{\rho}\langle w|w\rangle_{\sigma}|,
 \end{equation}
which evidently 
satisfies the conditions in Eq.~\eqref{cond} in the space spanned by the purifications  $\mathcal{H}_{w}$.  

{Focussing on a driven many-body quantum system, the density matrix evolves in time, resulting in an evolution of their corresponding purifications in Eq.~\eqref{eq:ip}. The purifications thus evolve in time as, }
\begin{equation*}
|w(t)\rangle=\sum_{i=1}^{4}\sqrt{p_i(t)}|\psi_{i}(t)\rangle\otimes|\alpha_{i}^{\prime}\rangle.
\end{equation*}
\ {In the above, we have exploited the fact  that that evolution operator does not operate on the ancillary states.} Finally, we arrive at  the LO defined in terms of the purifications as
\begin{equation}
{\cal L}^I(t)=\langle w(0)|w(t)\rangle.
\label{eq_LO_IP}
\end{equation}
\ {In the subsequent discussion, we analyse the possibility of zeros of the  LOs as defined in Eqs.~\eqref{eq_LO_fid} and \eqref{eq_LO_IP} to detect the existence
of DQPTs.} It  is straightforward to verify that both of these LOs reduce to the correct pure state limit.

 \section{Dissipative evolution following a sudden quench}
 
 \label{sec_sudden}
 
  In this section, we consider a one-dimensional  TFIM, described by the Hamiltonian \ct{sachdev10,suzuki13},
  \begin{equation}\label{eq:1}
 H=-\sum_{n=1}^{L}\sigma_n^{x}\sigma_{n+1}^{x}-h \sum_{n=1}^{L}\sigma_{n}^{z},
 \end{equation}
where $\sigma_n$s are Pauli spin matrices residing on the lattice  site $n$,
 and  study the post quench non-equilibrium dissipative dynamics  following a sudden
  change of the transverse field $h$  from an initial value $h_i$ to a final value $h_f$. 
At time $t=0$, the system is prepared in  the  ground state  represented by the pure state density matrix  $\rho(0)=|\psi_0\rangle\langle\psi_0|$ corresponding to the initial field $h_i$.

  Employing a Jordan-Wigner (JW) transformation $\sigma_{n}^{-}=e^{i\pi\left(\sum_{m<n}c_m^\dagger c_m\right)}c_n$ from spin-$1/2$ to spinless fermions $c_n$,   the final Hamiltonian with field $h_f$  can be
  recast to the form
 \begin{equation}\label{eq:2}
 H=- \sum_{n=1}^L\left(c_n^\dagger c_{n+1}+c_n^\dagger c_{n+1}^\dagger +h.c\right)+2h_fc_n^\dagger c_n,
 \end{equation}
where  $L$ is the system size. 
Due to  the discrete translational invariance of the system, after a Fourier transform with a periodic boundary condition the Hamiltonian in Eq.~\eqref{eq:2} gets decoupled for each momentum mode
$k$ and  assumes the form,
 \begin{equation}\label{eq:6}
 H_k(h_f)=  \left(\begin{array}{cccc}
 
 \cos{k}+h_f&0&0& \sin{k}\\
 0&0&0&0\\
 0&0&0&0\\
 \sin{k}&0&0&-(\cos{k}+h_f)
 
 \end{array}\right),
 \end{equation} 
 in the $4 \times 4$ Hilbert space spanned by the  basis vectors $|1\rangle_k=|0_k,0_{-k}\rangle$ (both fermionic states unoccupied),$\quad|2\rangle_k=|1_k,0_{-k}\rangle,\quad|3\rangle_k=|0_k,1_{-k}\rangle,\quad|4\rangle_k=|1_k,1_{-k}\rangle$ (both fermoinic states occupied).

We here consider the situation where the TFIM is coupled to a Markovian dissipative bath and thus the dynamics of the system is  dictated by a Lindblad master equation which at time
 $t$ is given by \ct{breuer_book},
 \begin{equation}\label{eq:4}
 \frac{d\rho}{dt}=-i[H(h_f),\rho]+D\left[\rho\right],
 \end{equation}
 where the dissipator $D$ is,
 \begin{equation}
 D\left[\rho\right]=\kappa\sum_n L_n\rho\L_n^\dagger-\frac{1}{2}\left\{L_n^\dagger L_n,\rho\right\},
 \end{equation}
with the notation $\{A,B\}$ denoting the anti-commutator between operators A and B.
However,  throughout this
work,  the bath is simulated by a decaying (particle-loss) type Lindblad operator,
 \begin{equation}\label{eq:3}
 L_n=c_n,
 \end{equation}  
 such that $L_n$ stays local in the JW fermion space.  The most remarkable aspect of such a Lindbladian is that, when Fourier transformed into momentum space it does not couple different momenta and consequently, the evolution dictating the dynamics can be decoupled in momenta to the form \ct{fazio17},
 \begin{equation}\label{eq:lk}
 \frac{d\rho_k}{dt}=-i[H_k(h_f),\rho_k
 ]+D_k\left[\rho_k\right],
 \end{equation}
where, the dissipator $D_k[\rho_k]$  transforms into a form decoupled in momenta,\
\begin{equation}\label{eq:dissipator}
D_k[\rho_k]=\kappa\left[ L_k\rho_kL_k^\dagger-\frac{1}{2}\left \{L_k^\dagger L_k,\rho_k\right\}\right],
\end{equation}
where $L_k=c_k$. 
Therefore,  the time evolved density matrix $\rho(t)$ also remains decoupled for each momenta mode, i.e.,
 \begin{equation}\label{eq:5}
 \rho(t)=\bigotimes_k \rho_k(t).
 \end{equation}
 where $\rho_k(t)$ is obtained by numerically solving Eq.~\eqref{eq:lk} which are $4\times4$ linear coupled differential equations for each $k$ as elaborated in Appendix \ref{ap5}.
 Hence, the complete dynamics of the TFIM can be  analysed  by inspecting the spectrum of the time evolved density matrix:
 \begin{equation}\label{eq:rho}
 \rho_k(t)=\sum_{j=1}^{4} p_{j}^k(t)|\psi_{j}^k(t)\rangle\langle\psi_{j}^k(t)|
 \end{equation}
  where $p_j^k(t)$ and $|\psi_j^k(t)\rangle$ are the eigenvalues and eigenvectors of $\rho_k (t) $ respectively.

 \subsection{Fidelity induced Loschmidt overlap}
 
 \label{sec_sudden_fidelity}
 
In this section, we follow a sudden quenching protocol followed by a dissipative evolution of the system density matrix, generated by a similar decaying Lindbladian. Starting from an initial ground state density matrix  $\rho_k(0)$, the fidelity between the initial and the time evolved density matrix  given by Eq.~\eqref{eq:fidelity} can be evaluated 
for the mode $k$ to obtain the form,
\begin{equation}\label{eq:21}
{\cal L}^{F}_{k}(\rho_k(0),\rho_k(t))=\sum_{j=1}^{4}p_{j}^k(t)|\langle\psi_{1}^k(0)|\psi_{j}^k(t)\rangle|^{2}.
\end{equation} 
Referring to the spectral decomposition of $\rho_k(t)$ as in Eq.~\eqref{eq:rho},
we note that the system is initially prepared in the ground state so that for the initial density matrix $\rho_k(0)$, $p_{1}^k(0)=1$ and $p_{i}^k(0)=0$ $\forall$ $i \neq$ $1$ and the 
initial ground state  $|\psi_{1}^k(0)\rangle$ and the excited state (orthogonal to the ground state) $|\psi_{4}^{k}(0)\rangle$ are coherent superpositions of the states $|1\rangle_k$ and $|4\rangle_k$. Further, {solving the linear system of equations following Eq.~\eqref{eq:lk},
 it is observed that 
$\langle\psi_{1}^k(0)|\psi_{2}^k(t)\rangle=\langle\psi_{1}^k(0)|\psi_{3}^k(t)\rangle=0$} (Appendix \ref{ap5})  and 
 \begin{figure*}
 	\begin{center}
 		\subfigure[]{\label{a}}{\centering\includegraphics[width=0.35\textwidth,height=0.79\columnwidth]{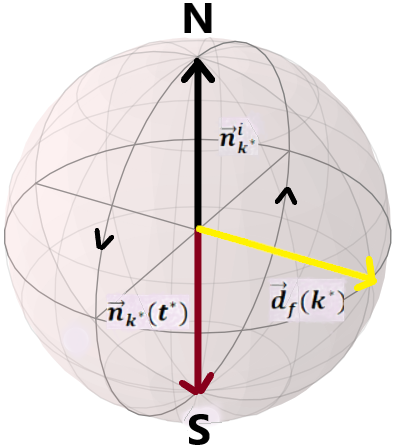}}
 		\hspace{3cm}\quad\subfigure[]{\label{b}}{\includegraphics[width=0.35\textwidth,height=0.79\columnwidth]{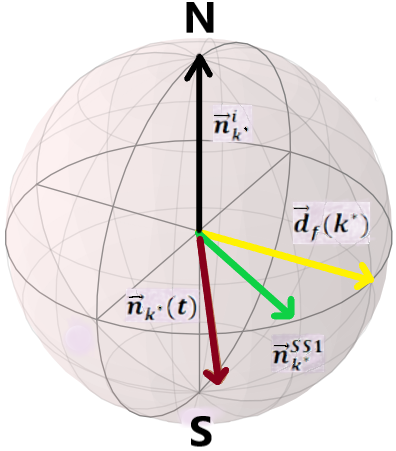}}
 		\quad\subfigure[]{\label{i}}{\includegraphics[width=0.45\textwidth,height=0.75\columnwidth]{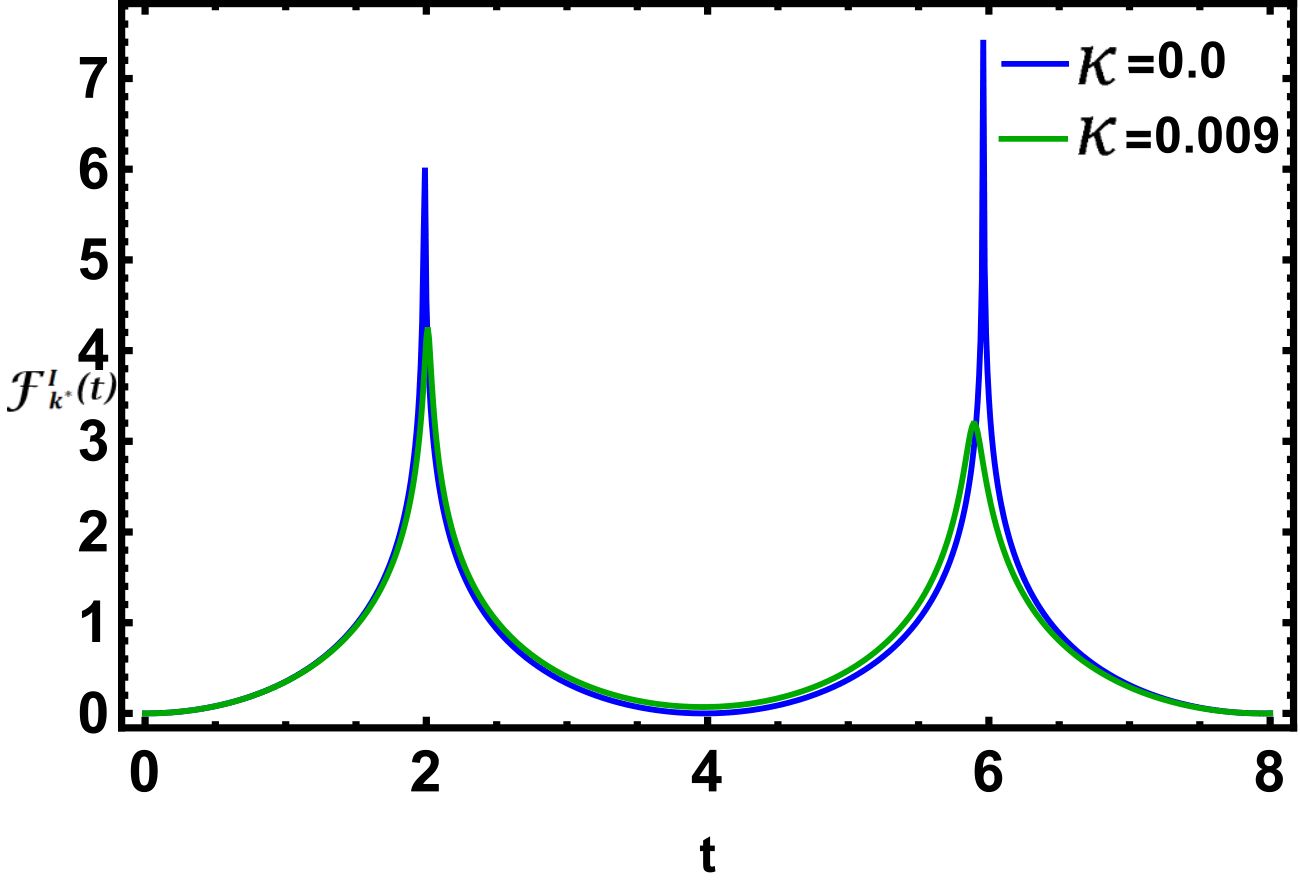}}
 		\quad\subfigure[]{\label{j}}{\includegraphics[width=0.45\textwidth,height=0.75\columnwidth]{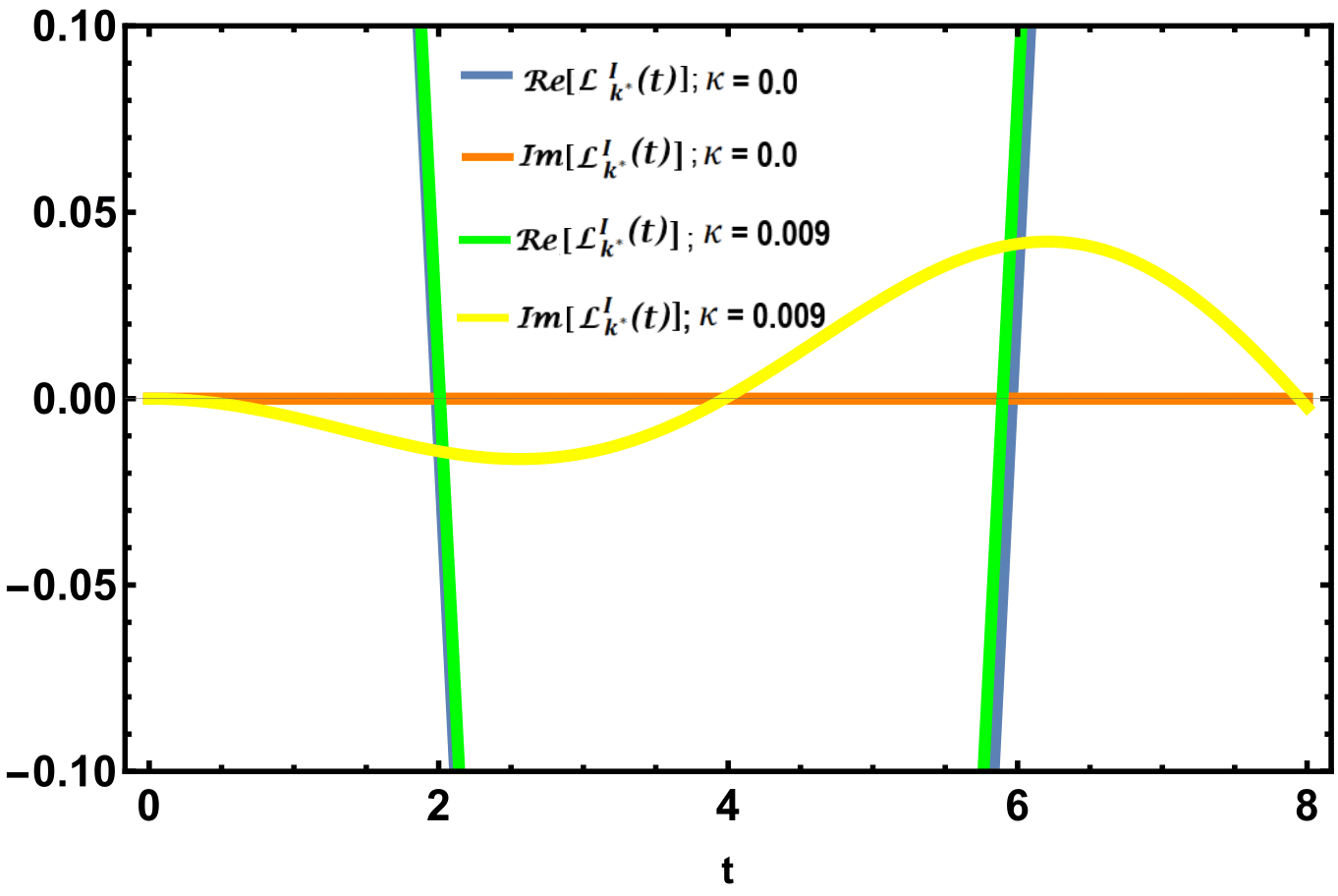}}

 		\caption{ (a) For the non-dissipative evolution ($\kappa=0$) following the sudden quench, the Bloch vector of a critical mode corresponding to the initial pure state $\vec{n}^i_{k^*}$ being perpendicular to the quenched Hamiltonian $\vec{d}_f(k^*)\cdot\vec{\sigma}$, rotates in a great circle, occasionally being antipodal to the initial direction signalling unitary DQPTs. (b) For a dissipative post quench evolution ($\kappa \neq 0$), on the other hand, the approach to  the dissipative steady state pulls the time evolved Bloch vector out of the plane containing $\vec{n}^i_{k^*}$ and its antipodal point thus taking the system away from dynamical criticality. 
 			(c) The free energy density ${\cal F}^{I}_{k^{*}}(t)$ for the critical mode $k^{*}=2.636$ shows non-analyticities corresponding to the critical times $t^{*}$s for dissipation strength, $\kappa=0.0$ and a quench from $h_i=0.5$ to $h_f=1.5$, whereas the non-analytic cusps smoothen out for any non-zero $\kappa$.
 			(d) The behaviour of the real and imaginary parts of the interferometric LO, ${\cal L}^{I}_{k^{*}}(t)$ is shown as a function of time following a sudden quench from $h_i=0.5$ to $h_f=1.5$. For $\kappa=0$, both the real and imaginary parts of the LO vanish at the critical instants of time. However, for a finite $\kappa$ ($\kappa$=0.009), the imaginary part of the LO deviates significantly from zero, and the deviation increases with increasing time.}
 	\end{center}
 \end{figure*}
 thus Eq.~\eqref{eq:21} reduces to
\begin{equation}\label{eq:22}
\begin{split}
{\cal L}^{F}_{k}(\rho_k(0),\rho_k(t))=p_{1}^k(t)|\langle\psi^k_{1}(0)|\psi^k_{1}(t)\rangle|^{2}\\
+p_{4}^k(t)|\langle\psi^k_{1}(0)|\psi^k_{4}(t)\rangle|^{2}.
\end{split}
\end{equation}
Let us emphasise that $p_1^k(t)$ and $p_4^k(t)$  can only vanish in  the steady state limit and furthermore the  overlaps $|\langle\psi_{1}^k(0)|\psi_{4}^k(t)\rangle|$ and $|\langle\psi_{1}^k(0)|\psi_{1}^k(t)\rangle|$ cannot simultaneously vanish. Therefore the LO defined in the fidelity approach in Eq. \eqref{eq_LO_fid} will not be able to capture DQPTs during the dissipative evolution as also observed in [\onlinecite{mera17}].

\begin{figure*}
	\begin{center}
		
		\quad\subfigure[]{\label{c}}{\includegraphics[width=0.35\textwidth,height=0.79\columnwidth]{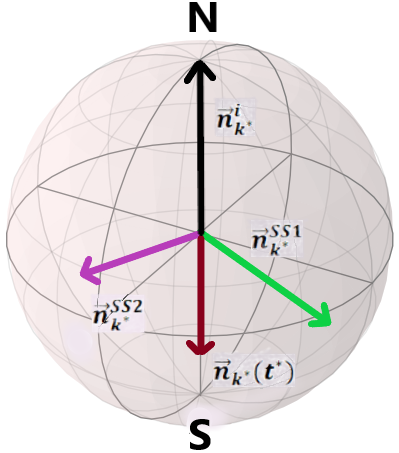}}
		\quad\subfigure[]{\label{d}}{\includegraphics[width=0.50\textwidth,height=0.75\columnwidth]{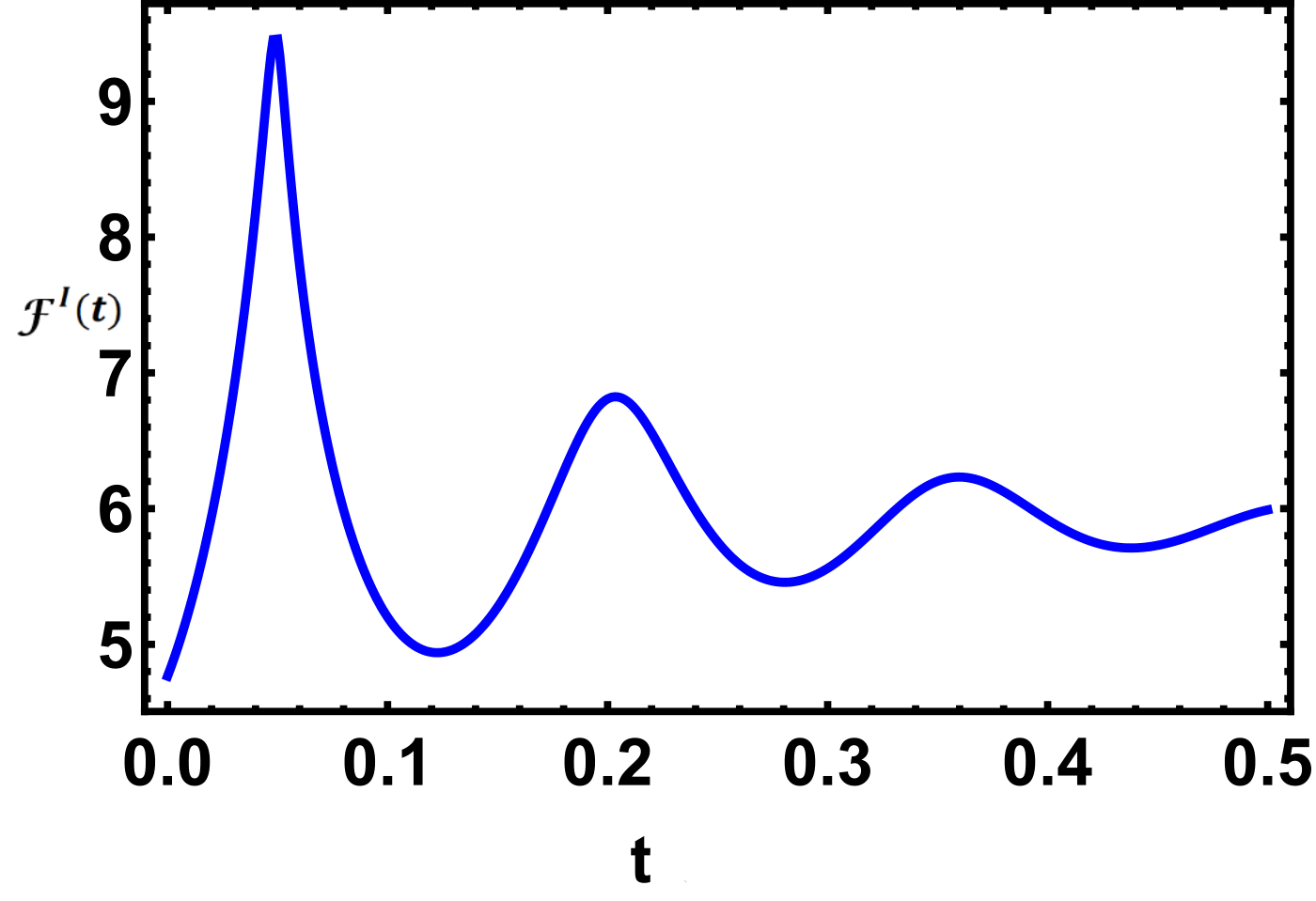}}
		\caption{ (a) A further quench after reaching the  first steady state $\vec{n}^{SS1}_{k^*}$,  drives the system out of equilibrium and the time evolved Bloch vector $\vec{n}_{k^*}(t)$ may become antipodal to the initial state in the approach to the new steady state $\vec{n}^{SS2}_{k^*}$, showing the possibility of dissipative inter steady state DQPTs in passage
		from one steady state to the other. 
			(b) Inter steady state dissipative DQPT in a TFIM following a sudden quench  of the external magnetic field to $h_f=20$ after reaching SS1 corresponding to $h_i=0$ and the dissipative coupling strength  $\kappa=5$. The dynamical free energy ${\cal F}^I(t)$ shows a non-analyticity at a critical instant at which the condition \eqref{eq:adqpt} is exactly satisfied for $k^{*}=\frac{\pi}{2}$.  We reiterate that the time $t$ is measured from the instant when a quench of the field  is applied on the steady state SS1.}
	\end{center}
\end{figure*}

\subsection{Interferometric phase induced Loschmidt overlap}

\label{sec_sudden_interfero}

For a driven quantum system, the time evolution of the density matrix $\rho_k(0)$ results in an evolution of their corresponding purifications  defined in Eq.~\eqref{eq:ip}. The purifications as defined in Eq.~\eqref{eq:ip} evolve in time as, 
\begin{equation*}
|w_k(t)\rangle=\sum_{i=1}^{4}\sqrt{p_i^k(t)}|\psi_i^k(t)\rangle\otimes|\alpha_{i}^{\prime}\rangle.
\end{equation*}
Since  the ancillary states do not evolve in time during evolution of the system,
the LO in terms of the purification assumes the form:
\begin{equation}\label{eq:25}
{\cal L}^I_{k}(t)=\langle w_{k}(0)|w_{k}(t)\rangle.
\end{equation}
 Also, a dynamical counterpart of the equilibrium free energy density may be defined in terms of the LO as,
 \begin{equation}\label{eq:dfe}
 	{\cal F}^I(t)=-\mathrm{Re} \left[\int_0^\pi \log{\left[{\cal L}^I_{k}(t)\right]}dk \right].
 \end{equation} 
The corresponding purifications at the initial and at a later time for the above quenched dissipative evolution are respectively considered as,
\begin{equation}\label{eq:23}
~|w_{k}(0)\rangle=\sum_{i=1}^{4}\sqrt{p_{i}^k(0)}|\psi_{i}^k(0)\rangle\otimes|\alpha_{i}\rangle
\end{equation}
and
\begin{equation}\label{eq:24}
\begin{split}
|w_{k}(t)\rangle=\sum_{i=2,3}\sqrt{p_{i}^k(t)}|\psi_{i}^k(t)\rangle\otimes|\alpha_{i}\rangle\\
~+~
\sqrt{p_{1}^k(t)}|\psi_{1}^k(t)\rangle\otimes|\alpha_{1}\rangle
+\sqrt{p_{4}^k(t)}|\psi_{4}^k(t)\rangle\otimes|\alpha_{4}\rangle,\\
\end{split}
\end{equation}
where $|\psi_{2}^k(t)\rangle=|\psi_{2}^k(0)\rangle$, $|\psi_{3}^k(t)\rangle=|\psi_{3}^k(0)\rangle$ (up to a phase) and \quad $|\psi_{1}^k(t)\rangle$ and $|\psi_{4}^k(t)\rangle$ are linear combinations of $|\psi_{1}^k(0)\rangle$ and $|\psi_{4}^k(0)\rangle$ (i.e., those of $|1\rangle_k$ and $|4\rangle_k$), under the chosen  Lindbladian (see Appendix~\ref{ap5}). Starting from the ground state of the initial Hamiltonian (a pure state) such that $p_{2}^k(0)=p_{3}^k(0)=0$, we study the dissipative evolution following a sudden quench in the external magnetic field.
In terms of the eigenvectors of the density matrix, the LO as defined in Eq.~\eqref{eq:25} may then be expanded to the form,
\begin{equation}\label{eq:26}
\begin{split}
\langle w_{k}(0)|w_{k}(t)\rangle=
\sqrt{p_{1}^k(0)}\sqrt{p_{1}^k(t)}\langle\psi_{1}^k(0)|\psi_{1}^k(t)\rangle +\\ \sqrt{p_{4}^k(0)}\sqrt{p_{4}^k(t)}\langle\psi_{4}^k(0)|\psi_{4}^k(t)\rangle.
	\end{split}
\end{equation}
We consider a relevant set of conditions which can give rise to DQPTs according to the LO defined in the interferometric approach, 
\begin{equation}\label{eq:27}
	\begin{split}
	\langle\psi_{1}^k(0)|\psi_{1}^k(t)\rangle=\langle\psi_{4}^k(0)|\psi_{4}^k(t)\rangle=0
	\end{split}.
\end{equation}
We note that DQPTs may also arise for certain isolated points in the parameter space that are not covered by the above conditions. The possibility of such occurrences of DQPTs being very rare, have not been studied further. 
Since the states $|\psi_{1}^k(t)\rangle$ and $|\psi_{4}^k(t)\rangle$ lie completely in the subspace $\{|1\rangle_k, |4\rangle_k\}$, it is sufficient to study the system density matrix in the subspace $\{|1\rangle_k,|4\rangle_k\}$, in the $2\times2$ form
\begin{equation}\label{eq:28}
\rho^{\prime}_{k}(t)=\frac{1}{2}(C_{k}(t)\mathbb{I}+\vec{n}_{k}(t)\cdot\vec{\sigma}),
\end{equation}
where $C_{k}(t)=p_{1}^{k}(t)+p_{4}^{k}(t)$ and $\mathbb{I}$ is $2 \times 2$ identity matrix.

The conditions for DQPTs \eqref{eq:27} may be recast in terms of the initial and time evolved Bloch vectors as (Appendix \ref{ap1})
\begin{equation}\label{eq:29}
	\hat{n}_{k^{*}}(0)\cdot\hat{n}_{k^{*}}(t^*)= -1,
\end{equation}
for a critical mode $k^{*}$. That is, $\vec{n}_{k^{*}}(t^*)$ and $\vec{n}_{k^{*}}(0)$ are antipodal on the Bloch sphere (see Fig.~\ref{a}). This condition is similar to the case of DQPTs in completely unitary evolution where the quenched Hamiltonian generates a rotation of the Bloch vector $\vec{n}_{k^{*}}(t)$ on the Bloch sphere such that it satisfies the condition \eqref{eq:29} at certain critical instants \ct{heyl13}. The motion being periodic, the dynamical free energy density as defined in Eq.~\eqref{eq:dfe} becomes non-analytic periodically at the critical instants, signalling DQPTs. Despite, the similarities in the case of a dissipative evolution the Bloch vector is not restricted to lie on the Bloch sphere.

\begin{figure*}
	\subfigure[]{\label{e}}{\centering\includegraphics[width=0.48\textwidth,height=0.75\columnwidth]{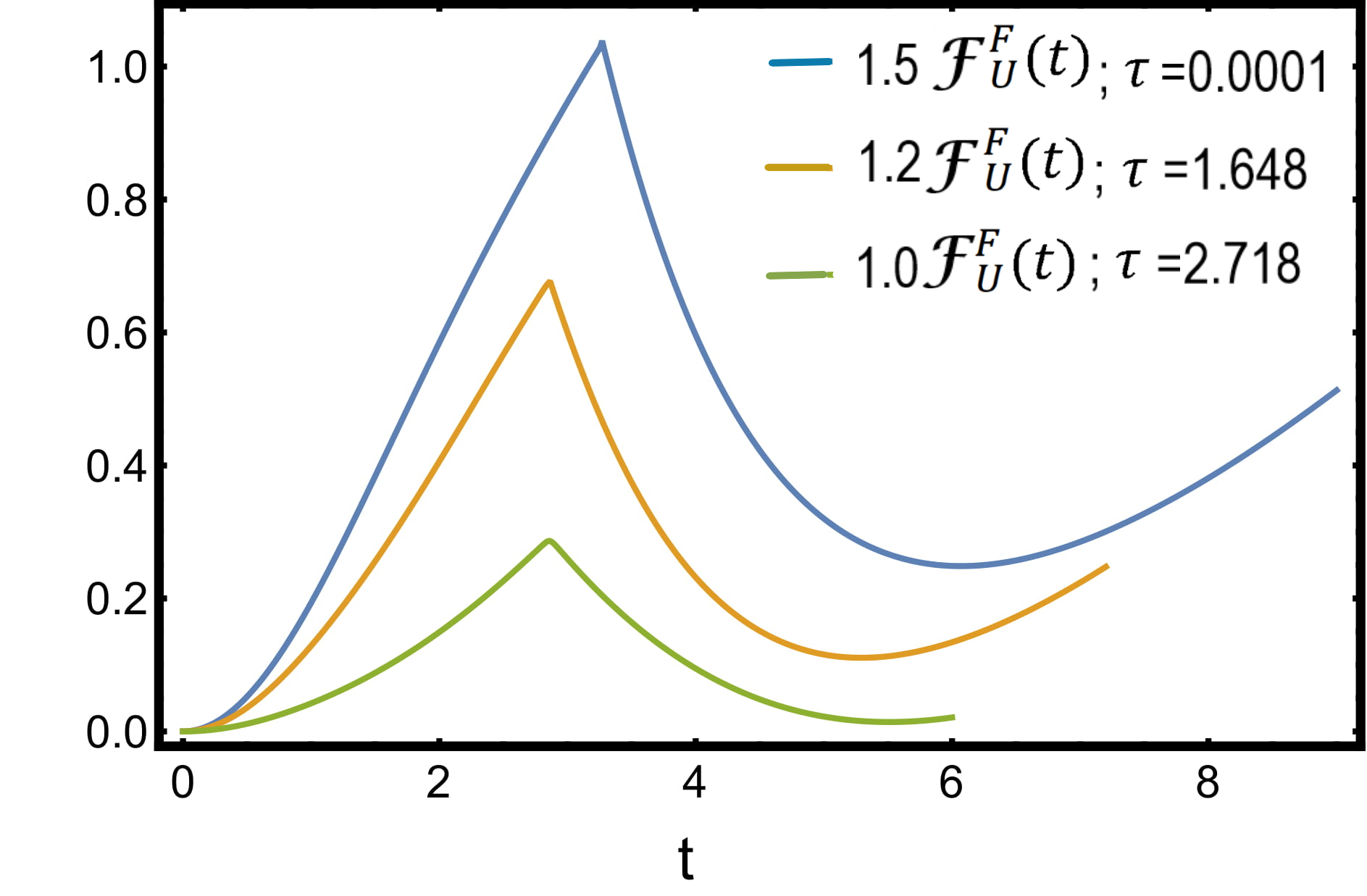}}
	\subfigure[]{\label{f}}{\includegraphics[width=0.48\textwidth,height=0.75\columnwidth]{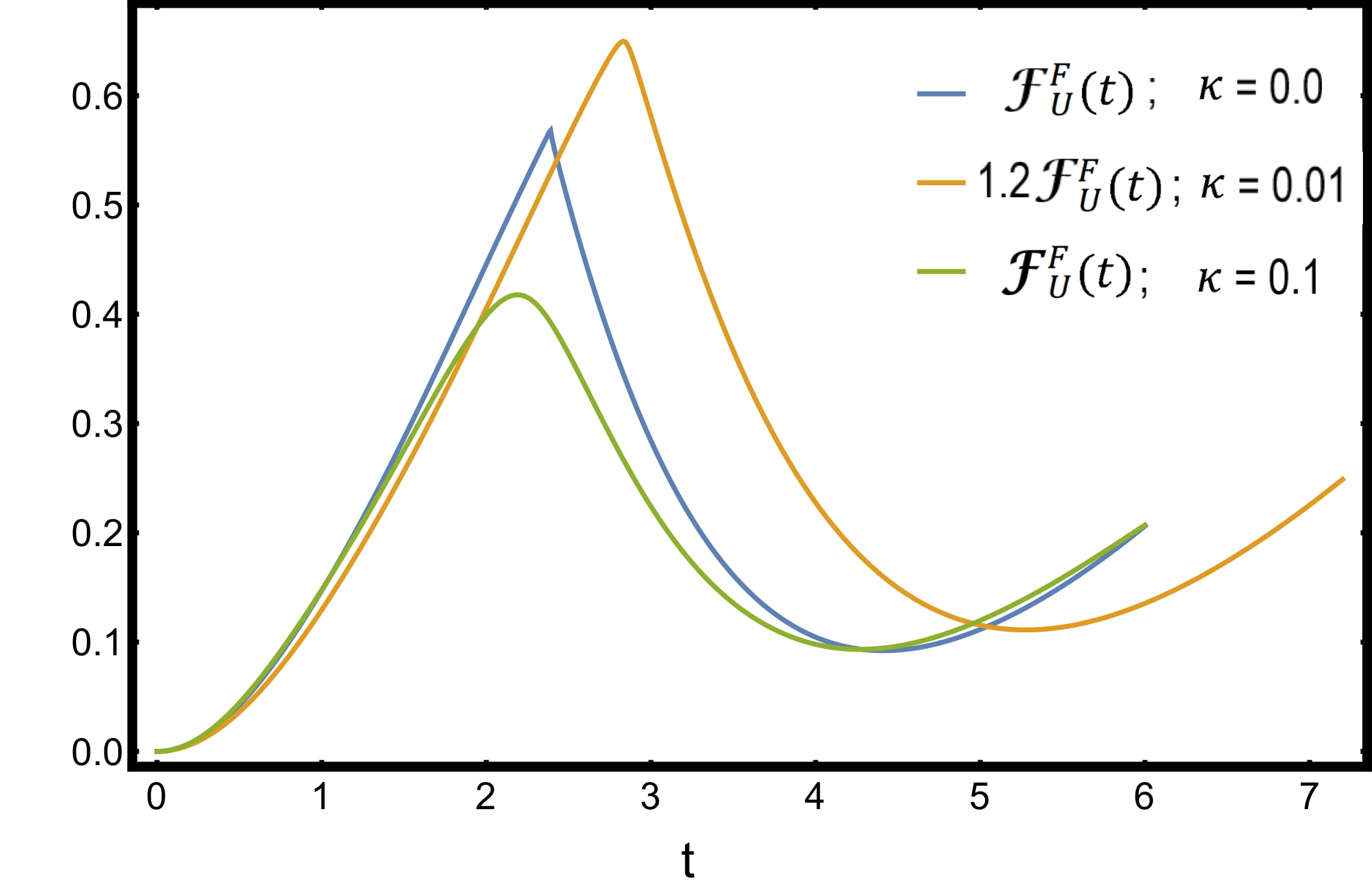}}
	\caption{(a) The  smoothening of non-analyticities in the dynamical free energy density given by Eq.~\eqref{eq:f1}, with changing quenching time $\tau$ for a fixed value of dissipation strength $\kappa = 0.001$. The ramping rate $\tau=0.0001$ is so chosen that the dynamics lies  in the Kibble-Zurek dominated  regime of the residual energy \ct{fazio17} and has an apparent  non analyticity (though not a true DQPT) in the dynamical free energy. 	(b) The  smoothening of non-analyticities in the dynamical free energy density given by Eq.~\eqref{eq:f1}, with dissipation strength $\kappa$  for a fixed value of ramping time $\tau = 2.718$. The case $\kappa=0$ signifies no dissipation and has a clear non analyticity in the dynamical free energy density.  In both the cases, ${\cal F}_{U}^F(t)$s have been scaled differently for clear distinguishability of the peaks.}
\end{figure*}

However for a Lindbladian evolution there necessarily exists a steady state density matrix such that,
\begin{equation}\label{eq:30}
\lim\limits_{t\rightarrow\infty}\rho^{\prime}_k \to \rho_k^{\prime,SS} \implies \lim\limits_{t\rightarrow\infty}\vec{n}_{k}(t) \to \vec{n}_{k}^{SS}~{\rm and}~C_k(t) \to C_k^{SS}.
\end{equation}   
The existence of the steady state  restricts any periodic motion of $\vec{n}_{k}(t)$ within the Bloch sphere (Fig.~\ref{b}). The approach to the steady state always moves the system away from the dynamical critical points resulting in a  disappearance of the DQPTs
for any non-zero value of $\kappa$. We note that, greater the deviation of unitary critical time, $t^{*}_{j}$ from the dissipative time scale $(\kappa^{-1})$, more rapid is the deviation of LO from zero (see Fig.~\ref{i},\ref{j}).

\subsubsection{Inter-steady state DQPTs}

\label{sec_sudden_steady_state}

In the previous section we illustrated, that when the system is allowed to evolve in the presence of dissipation governed by the above Lindbladian following a sudden quench,  the system reaches a dissipative steady state SS1 ($\rho_{SS1}$) and the dynamical nature of the density matrix is lost. In this section, we consider a further quench of the transverse field $(h)$ which eventually drives the system to a new steady state SS2 ($\rho_{SS2}$) through a dissipative dynamics. We show that interestingly the occurrence of DQPTs during the evolution from the steady state SS1 to the steady state SS2 (see Fig.~\ref{c}) is indeed possible. \\

Starting from an initial pure state characterised by a $2 \times 2$ density matrix $\rho^{\prime}_{0}$, the dissipative evolution following a sudden quench drives the system to the steady state  SS1.  The transverse field is then again suddenly changed to a new value which further introduces dynamics in the system and accidental DQPTs are possible if at any time $t^{*}$ there exists a critical mode $k^{*}$ such that 
\begin{equation}\label{eq:adqpt}
	\hat{n}_{k^{*}}(0)\cdot\hat{n}_{k^{*}}(t^{*})= -1,
\end{equation}
where $\hat{n}_{k^{*}}(0)$ now is the Bloch vector representing the initial density matrix $\rho^{\prime}_0$;  $\hat{n}_{k^{*}}(t^{*})$, on the other hand, is the Bloch vector representing the density matrix at time $t^{*}$ measured from the instant when the second quench is performed on the SS1. Thus the system may show a transition between two steady states separated by a DQPT associated with non-analyticities in the dynamical free energy density defined in Eq.~\eqref{eq:dfe} at the critical instant  $t^{*}$ for
some values of quenching amplitude and the dissipation strength (see Fig.~\ref{d}). This observation indeed widens the horizon of the occurrence of DQPTs  in open quantum systems and their experimental realisations as discussed below. These inter-state DQPTs are also possible for a dissipative evolution following a dissipative slow ramping state preparation discussed in the subsequent section.

\section{Unitary evolution following a state preparation with dissipative linear ramping} 

\label{sec_unitary}

\subsection{Preparation of the pre-quench state through ramping}

\label{sec_unitary_preparation}

In this section, we shall prepare the initial density matrix using a dissipative slow ramping and probe the occurrence of DQPTs in the subsequent
\textit{unitary} evolution. To achieve this,
we now adopt a dissipative state preparation protocol through a slow ramping in the presence of the bath in  such a way that the transverse field is varied across a QCP  following the protocol: \begin{equation}\label{eq:slow}
	h(t_1)=h_{i}+(h_{f}-h_{i})\frac{t_1}{\tau},
\end{equation}
 where $h_{i}$ and $h_{f}$ are the initial and final fields, $\tau$ being the total quenching time (rate). In this process, the system gets prepared in the  density matrix $\rho(\tau)$; setting the origin of time at $t_1=\tau$, we then calculate the LO due to the  {unitary} temporal evolution generated by the time-independent final Hamiltonian $H(h_f)$ and explore the possibilities of DQPTs.

\subsection{Fidelity induced Loschmidt overlap}

\label{sec_unitary_fidelity}

 Following the slow quench, the density matrix evolves unitarily under $U_k(t)= \exp(-iH_k(h_f)t)$; thus,  $\rho_k(t)=U_k(t)\rho_k(0)U_k^\dagger(t)$ ; where $\rho_k(0)=\rho_{k}(\tau)$ describes the system just after the drive is over, i.e.,
at  time $t_1 =\tau$ and time $t$ is measured from $t_1=\tau$. The distance between $\rho_{k}(0)$ and $\rho_{k}(t)$ according to the fidelity defined in Eq.~\eqref{eq:fidelity} serves as a LO,
\begin{equation}\label{eq:flo}
{\cal L}^F_U(t)=\prod_k \mathrm{Tr}\left[\sqrt{\sqrt{\rho_k(0)}\rho_k(t)\sqrt{\rho_k(0)}}\right].
\end{equation}

\noindent We again  define the dynamical free energy density:
\begin{equation}\label{eq:f1}
{\cal F}_{U}^F(t)=-\mathrm{Re}\left[\int_0^\pi\log{\mathrm{Tr}\left[\sqrt{\sqrt{\rho_k(0)}\rho_k(t)\sqrt{\rho_k(0)}}\right]dk}\right].
\end{equation}
and probe the non-analyticities in the dynamical free energy density ${\cal F}_{U}^F$ signalling the real time DQPTs.\\
As the Hamiltonian which generates the time evolution of the system  acts only on the subspace $\{|1\rangle_k, |4\rangle_k\}$, the eigenvalues of the density matrix $\rho_k(0)$ i.e.,  $p_1^k, p_2^k, p_3^k, p_4^k$ remain invariant in time during the post-quench unitary dynamics. The Loschmidt overlap defined in Eq.~\eqref{eq:flo} for each $k$-mode is related to (see Appendix.~\ref{ap2}),

\begin{multline}\label{eq:13}
 \mathrm{Tr}\left[\sqrt{\sqrt{\rho_k(t)}\rho_k(0)\sqrt{\rho_k(t)}}\right]= p^k_2+p^k_3+\\
\sqrt{\left(\sum_{i=1,4,j=1,4}p_i^k p_j^k|\langle\psi_i^k(0)|U_k(t)|\psi_j^k(k)\rangle|^2
+2\prod_{i=1,4}p_i^k\right)}
.
\end{multline}

We again note that all the terms on the right hand side of Eq.~\eqref{eq:13} are positive, it does not vanish as all the $p_i^k$'s are nonzero, however small they may be. Since, the state preparation protocol is dissipative, the non-analyticities are wiped out and the Lindbladian evolution leads to an {\it initial mixed density matrix}.
Hence, starting from a mixed state, the LO constructed from the fidelity between density matrices does not indicate the occurrence of any DQPT (Fig.~\ref{e})
even when the subsequent dynamics is unitary. 

\subsection{Interferometric phase induced Loschmidt overlap}

\label{sec_unitary_interfero}

Simplifying Eq.~\eqref{eq:25}, we define the LO between two density matrices in this approach as ,
\begin{equation}\label{eq:14}
{\cal L}^I_{U,k}(t)=\mathrm{Tr}\left[\rho_k(0)U_k(t)]\right];~~{\rm and}~~{\cal L}_U^I(t)=\prod_{k} {\cal L}^I_{U,k}(t),
\end{equation}
with the associated dynamical free energy density:
\begin{equation}\label{eq:16}
{\cal F}^I_{U}(t)=-\mathrm{Re}\left[\int_0^\pi \log{\mathrm{Tr}\left[\rho_k(0)U_k(t)\right]}dk \right].
\end{equation}

\begin{figure*}
\subfigure[]{\label{g}}{\centering\includegraphics[width=0.45\textwidth,
height=0.65\columnwidth]{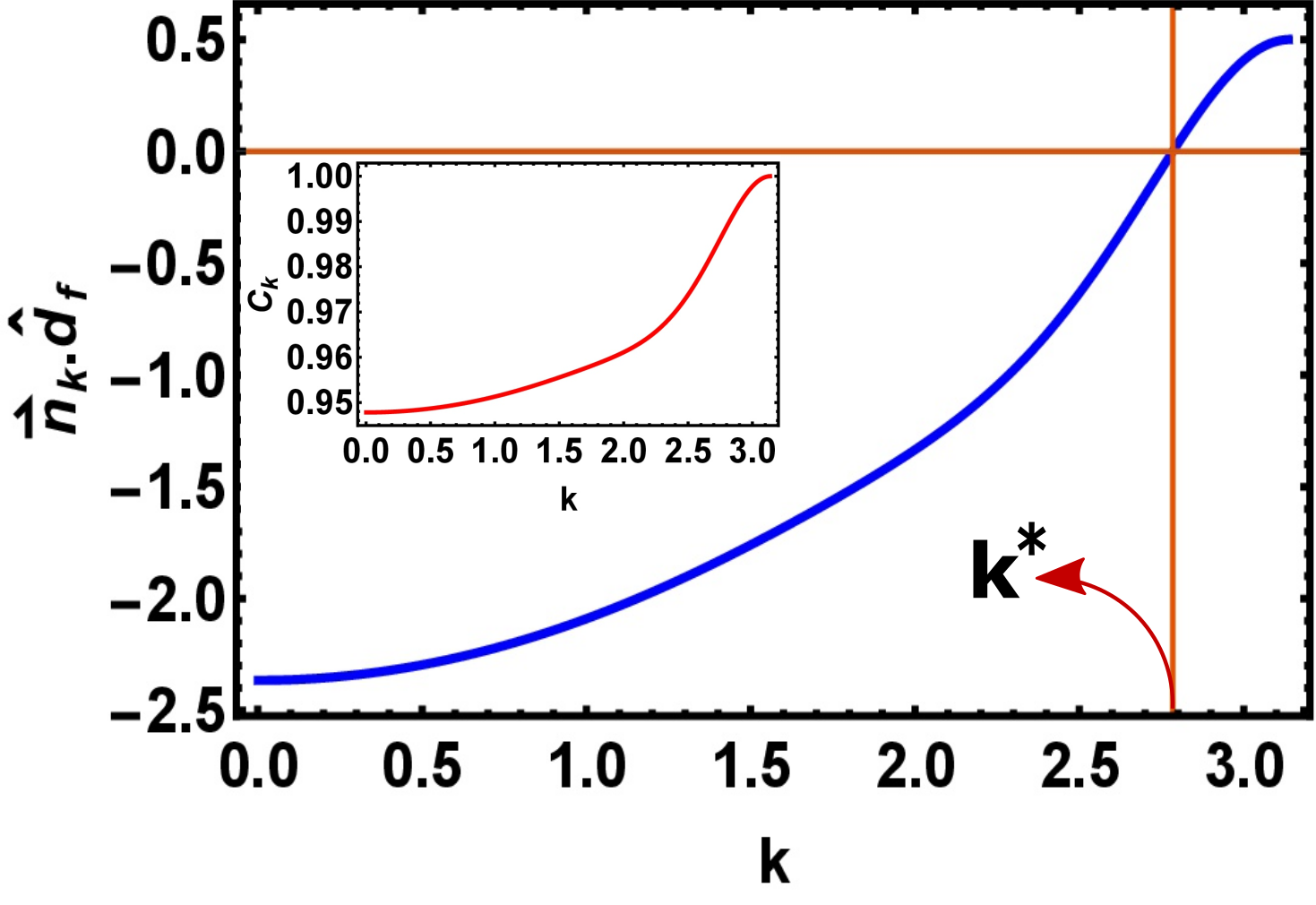}}
\quad\subfigure[]{\label{h}}{\includegraphics[width=0.45\textwidth,height=0.65\columnwidth]{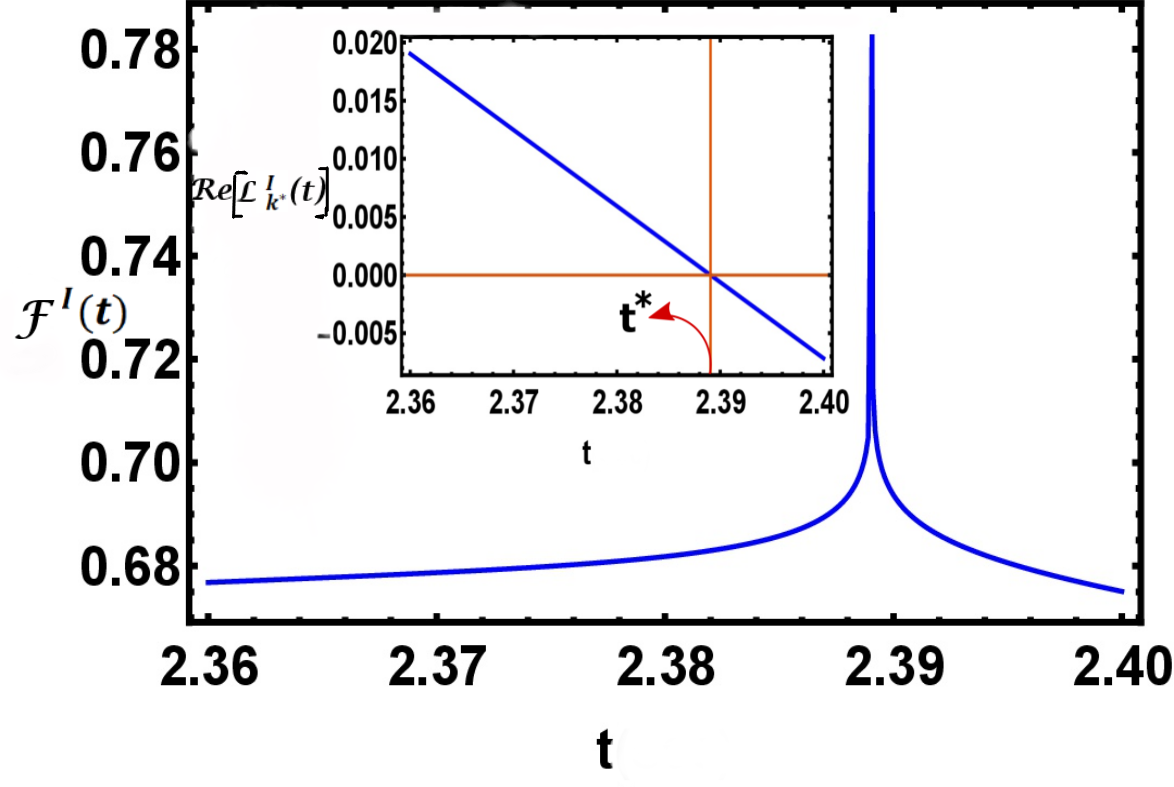}}
\caption{(a) The scalar product of the post-quench Bloch vector with the final Hamiltonian crosses zero at a critical $k^{*}$, hence satisfying the condition for criticality Eq.~\eqref{eq:abcd}.\quad Inset: The quantity $C_k$ is plotted as a function of $k$ revealing that it indeed satisfies the condition Eq.~\eqref{eq:36}, allowing the existence of critical times.\quad(b) The interferometric free energy density shows non-analyticities signalling DQPTs for a TFIM (system size  $L=500$) during an unitary evolution following a dissipative density matrix preparation. The initial mixed state is prepared by slowly ramping the external magnetic field $h$ from $h_i=0.5$ to $h_f=1.5$ across a QCP over a time span of $\tau=e$  with the  dissipation strength $\kappa=0.01$.\quad Inset: The real part of the interferometric LO for the critical momenta ($k^{*}$) vanishes at a critical time $t^{*}$ satisfying the condition for criticality Eq.~\eqref{eq:20}, thus signalling DQPTs.}
\end{figure*}

Unlike the LO constructed from the fidelity, $\mathrm{Re}[{\cal L}^{I}_{U}(t)]$ can assume both positive and negative values while  the former is a positive definite measure; this, in turn, renders the
possibility of observing DQPTs through the interferometric measure.\\
Recalling that the time-independent final Hamiltonian $H_f$ is diagonal in the subspace $ \{|2\rangle_k,|3\rangle_k \}$ we write an effective $2\times2$ Hamiltonian $H_{eff}=\vec{d}_{f}(k).\vec{\sigma}$ in the subspace $\{|1\rangle_k,|4\rangle_k\}$ where $\vec{d}_{f}(k)= (\sin{k},0,\cos{k}+h_{f})$. The eigenvalues of $H_{eff}$ are $\pm\epsilon^{f}_{k}=\pm |\vec{d}_{f}(k)|$. Also the density matrix just after the quench for each $k$ mode is observed to be diagonal in the subspace $\{|2\rangle_k,|3\rangle_k\}$ implying that the post-quench dynamics of the system is effectively same as the unitary evolution of a two-level system connecting the states $\{|1\rangle_k,|4\rangle_k\}$. Let us
again make recourse to the $2\times2$ matrix,
\begin{equation}\label{eq:17}
\rho_{k}^{\prime}(0)=\frac{1}{2}(C_{k}\mathbb{I}+\vec{n}_{k}\cdot\vec{\sigma}),                                                                                                                                                     
\end{equation}
where $C_{k}=p^{k}_1 + p^{k}_4$ is time independent.
The critical times and critical momenta of the DQPTs are determined by the zeros of the LO, i.e., 
\begin{equation}\label{eq:18}
{\cal L}^{I}_{U,k^{*}}(t^{*})=\mathrm{Tr}\left[\rho^{\prime}_{k^{*}}(0)e^{-\mathrm{i}H_{eff}(k^{*})t^{*}}\right]+(1-C_{k^{*}})=0.
\end{equation}
The above equation simplifies to (see Appendix.~\ref{ap2}),
\begin{equation}\label{eq:19}
C_{k^{*}}\cos{(\epsilon_{k^{*}}^{f}t^{*})}-\mathrm{i}\sin{(\epsilon_{k^{*}}^{f}t^{*})}\vec{n}_{k^{*}}\cdot\hat{d}_{f}(k^{*})+(1-C_{k^{*}})=0.
\end{equation}
Thus the critical momenta are obtained by solving the equation, 
\begin{equation}\label{eq:abcd}
\vec{n}_{k^{*}}\cdot\hat{d}_{f}(k^{*})=0	
\end{equation}
 and the critical times are solutions of,
\begin{equation}\label{eq:20}
\cos{(\epsilon_{k^{*}}^{f}t^{*})}=1-\frac{1}{C_{k^{*}}}.
\end{equation}
The expression for the critical times, i.e., Eq.~\eqref{eq:20} clearly imposes a bound on the coefficient $C_{k^{*}}$ for the existence of DQPTs which is
\begin{equation}\label{eq:36} 
C_{k^{*}}\geq \frac{1}{2}.
\end{equation}
What is remarkable is that, the bound on $C_{k^{*}}$ imposes a condition on the dissipation strength $\kappa$ and the duration of the ramping $\tau$ which defines a specific region in the $\kappa-\tau$ plane which supports the existence of DQPTs following a dissipative state preparation (Fig.~\ref{h}). We further note that no such bounds exist in the case when the bath is absent during the state preparation \ct{sharma16}. This constraint relation
is unique to the present work where one must take into consideration the 4-level nature of the  decoupled Hamiltonian  for each momenta when the unitary dynamics
sets in. This constraint arises because the Hamiltonian generating the unitary dynamics acts only on the subspace spanned by $\{|1\rangle_k, |4\rangle_k\}$.\\
 In analogy with equilibrium phase transitions, we can now consider the LO as a dynamical order parameter which can distinguish between the dynamical phases in the $\kappa-\tau$ plane. When $C_{k^*}\geq 1/2$, $k^*$s can always be found for which the LO is zero (gapless). On the other hand, the LO remains non-zero (gapped) for any $k$ when $C_{k}<1/2$. Thus, we find the semblance  of a phase transition from a gapless LO to a gapped LO phase separated by critical boundary in the $\kappa-\tau$ plane parametrized by critical $\kappa_c$ and $\tau_c$.  Therefore, tuning $\kappa$ and $\tau$, one can drive the system
form one phase to the other.

\section{Experimental possibilities}

\label{sec_expt}

In a recent experiment \ct{jurcevic16}, DQPTs have been directly observed in interacting many body TFIM simulated in a trapped ion system. In the reported experiment, the ground states of the initial Hamiltonian span a degenerate manifold. For the case of degenerate ground states, the pure state LO is redefined to be the return probability of the time evolved state to the complete ground state manifold. The rate function (or the return probability) measured by performing repetitive projective measurements on the time evolved state shows non-analyticities in the thermodynamic limit when the time evolved state reaches the perfectly symmetric combination in the ground state manifold.
Defining a similar generalized LO in the interferometric approach, one may observe non-analyticities in the rate function during the transition of the system between different steady states. This possibility is higher for the chosen type of bath, as the ground state manifold of the initial Hamiltonian is invariant in time under the action of the Lindbladian defined in Eq.~\eqref{eq:3}. In recent experiments \ct{thekkadath16}  using combinations of weak and strong measurements, it has been possible to measure all the elements of a mixed state density matrix in a chosen basis. Similar experimental ventures will make the tomography of many-body mixed state density matrices possible which in turn will result in the possible measurement of the dissipative rate function showing dissipative DQPTs. Although the chosen bath being local in the JW fermions,  is  difficult to realise for spin chains, implementation for a similar bath would  indeed be plausible  in the one dimensional Kitaev chain. 

\section{Concluding comments}

\label{sec_conclusion}

In this work, we have  considered  two different versions of LOs, one extracted from the fidelity and other from the interferometric phase,  both of which reduce to the exact form of the LO  in the pure state limit. The  initial aim here is
   to explore  DQPTs and study a dissipative evolution following a sudden quench in the Hamiltonian.  The fidelity induced Loschmidt overlap is found to be inappropriate to capture any DQPT for a dissipative evolution. As for the interferometric phase induced LO, the existence of a dissipative steady state, takes the system  away from any dynamical criticality until finally the dynamics  freezes into the steady state. However, a second quench after the system has reached the steady state, drives the system towards a different steady state and  remarkably, dissipative DQPTs are indeed possible during the inter steady state dynamics as shown in Fig.~\ref{d}.
 This possibility of inter-steady state DQPTs is uniquely unravelled in the interferometric phase approach and has not been reported earlier.\\

In contrast to the dissipative post quench evolution, we also study the situation of unitary post quench dynamics following a dissipative state preparation by ramping the external magnetic field of a TFIM across a QCP;  the effect of two competing time scales, the quenching rate $\tau$ and $\kappa^{-1}$, on  the subsequent unitary dynamics when the bath is removed, is
explored in the context of  DQPTs. Fig.3 presents results obtained in the fidelity approach where we show the smoothening of DQPTs as reflected  in  the associated dynamical free energy density as in Eq.~\eqref{eq:13}. On the contrary, considering the dynamical free energy density defined in the interferometric phase approach one obtains a condition Eq.~\eqref{eq:36} which can be identified with a bound on the population of the subspace spanned by $\{|1\rangle_k, |4\rangle_k\}$. We reiterate that  this condition is an artefact of dissipative state preparation and is unique to the present study. This constraint relation also marks the boundary in the $\kappa-\tau$ plane between two dynamically inequivalent phases, in one the LO vanishes for the critical mode
$k^*$, while in the other no critical mode can exist and therefore the LO never vanishes.  Thus, we find a phase transition like situation from a gapless LO to a gapped LO phase separated by the critical boundary. 
We conclude with the note that the results presented in this paper would hold true for any generic bath that is local in JW space in any JW solvable system. \\

\section*{Acknowledgement}
AD acknowledges SERB, DST, New Delhi and SB acknowledges CSIR, New Delhi for financial support.
\appendix

\section{\textbf{Solving for the time evolved density matrix}}\label{ap5}

We shall elaborate on the set of $4\times4$ linear coupled first order differential equations determining the evolution of the time evolved density matrix for a particular $k$ in the basis spanned by the states $\{|1\rangle_k, |2\rangle_k,|3\rangle_k,|4\rangle_k\}$. Expanding the Lindblad master equation in Eq.~\eqref{eq:lk}, we find:

\begin{eqnarray}
\dot{\rho}^k_{11}=ig\rho^k_{14}-ig\rho^k_{41}+\kappa\rho^k_{22}+\kappa\rho^k_{33}\label{eq:s1}\\
\dot{\rho}^k_{14}=-\kappa\rho^k_{14}+ig\rho^k_{11}-ig\rho^k_{44}-2ic\rho^k_{14}\label{eq:s2}\\
\dot{\rho}^k_{22}=-\kappa\rho^k_{22}+\kappa\rho^k_{44}\label{eq:s3}\\
\dot{\rho}^k_{33}=-\kappa\rho^k_{33}+\kappa \rho^k_{44}\label{eq:s4}\\
\dot{\rho}^k_{44}=-ig\rho^k_{14}+ig\rho^k_{41}-2\kappa\rho^k_{44}\label{eq:s5}\\
\dot{\rho}^k_{23}=-\kappa\rho^k_{23}\label{eq:s6}\\
\dot{\rho}^k_{12}=-\frac{\kappa}{2}\rho^k_{12}+\kappa\rho^k_{34}-ig\rho^k_{42}-ic\rho^k_{12}\label{eq:s7}\\
\dot{\rho}^k_{13}=-\frac{\kappa}{2}\rho^k_{13}+\kappa\rho^k_{24}-ig\rho^k_{43}-ic\rho^k_{13}\label{eq:s8}\\
\dot{\rho}^k_{24}=-\frac{3\kappa}{2}\rho^k_{24}+ig\rho^k_{21}-ic\rho^k_{24}\label{eq:s9}\\
\dot{\rho}^k_{34}=-\frac{3\kappa}{2}\rho^k_{34}+ig\rho^k_{31}-ic\rho^k_{34}\label{eq:s10},
\end{eqnarray}
where $\dot{\rho}^k_{mn}={d\rho^k_{mn}}/{dt}$; $g=\sin{k}$ and $c=\cos{k}+h_f$, are parameters of the Hamiltonian.
The above $10$ equations along with complex conjugates of the off-diagonal terms  constitute the dynamics of the complete system for the mode $k$.  
The coupled equations ~\eqref{eq:s6} to ~\eqref{eq:s10} as well as  their complex conjugates are evidently independent  of the terms $\rho^k_{14} ~,~\rho^k_{41}$ and the diagonal terms of $\rho_k(t)$. We note that all off-diagonal terms of the density matrix except $\rho^k_{14} ~\text{and}~\rho^k_{41}$ are identically zero initially at $t=0$.  Therefore, the above equations imply that these off-diagonal components will remain fixed to zero throughout the temporal evolution of the density matrix.\\

Thus, the only non-zero components of $\rho_k(t)$ to be taken into consideration are $\rho^k_{11}(t), \rho^k_{14}(t), \rho^k_{41}(t), \rho^k_{44}(t), \rho^k_{22}(t) ~\text{and}~ \rho^k_{33}(t)$.
Also, the vanishing of $\rho^k_{2i}~\forall~i\neq 2~\text{and}~\rho^k_{3j}~\forall~j\neq 3$, clearly indicates that the states $|2\rangle_k$ and $|3\rangle_k$ can differ from the eigenstates of the time evolved density matrix only by a phase. Therefore, identifying the eigenstates of $\rho_k(t)$  with $\{|\psi^k_i(t)\rangle\}_{i=1}^4$, it is clear that 
$|\psi^k_1(t)\rangle~\text{and}~|\psi^k_4(t)\rangle$ 
are time dependent linear incoherent superpositions of $|1\rangle_k$ and $|4\rangle_k$ and,
\begin{multline}\label{eq:s11}
|2\rangle_k=|\psi^k_2(0)\rangle=|\psi^k_2(t)\rangle~\text{and}~\\
|3\rangle_k=|\psi^k_3(0)\rangle=|\psi^k_3(t)\rangle ,
\end{multline}
upto a phase. To be more specific, levels $|2\rangle_k$ and $|3\rangle_k$ only evolve up to a phase although the population of these levels change in the course of dynamics. Also, from the orthogonality of eigenstates,
\begin{equation}\label{eq:s12}
\begin{split}
\langle\psi_{1}^k(0)|\psi_{2}^k(t)\rangle=\langle\psi_{1}^k(0)|\psi_{3}^k(t)\rangle=0.
\end{split}
\end{equation}

\section{\textbf{Condition of orthogonality of two Bloch vectors}}\label{ap1}
Considering two $2 \times 2$ density matrices,
\begin{equation}\label{C1}
  \begin{split}
  \rho_1=\frac{1}{2}(\mathbb{I}+\vec{n}_1\cdot\vec{\sigma}),\\
  \rho_2=\frac{1}{2}(\mathbb{I}+\vec{n}_2\cdot\vec{\sigma}),
  \end{split}
\end{equation}
with corresponding eigen vectors $|\psi_{1}\rangle \text{and} |\psi_{2}\rangle$ such that,
\begin{equation}\label{C2}
\begin{split}
\vec{n}_{1}\cdot\vec{\sigma}|\psi_{1}\rangle=|\vec{n}_{1}||\psi_{1}\rangle,\\
\vec{n}_{2}\cdot\vec{\sigma}|\psi_{2}\rangle=|\vec{n}_{2}||\psi_{2}\rangle.
\end{split}
\end{equation}
The overlap between $|\psi_{1}\rangle~\text{and}~|\psi_{2}\rangle$ using Eq.~\eqref{C2} is shown to be
\begin{multline}\label{C3}
\langle\psi_{2}|\psi_{1}\rangle=\langle\psi_{2}|\frac{\vec{n}_{2}\cdot\vec{\sigma}}{|\vec{n}_{2}|}\frac{\vec{n}_{1}\cdot\vec{\sigma}}{|\vec{n}_{1}|}|\psi_{1}\rangle\\
=\langle\psi_{2}|\frac{\vec{n}_{2}\cdot\vec{n}_{1}}{|\vec{n}_{2}||\vec{n}_{1}|}+i\frac{(\vec{n}_{2}\times\vec{n}_{1})\cdot\vec{\sigma}}{|\vec{n}_{2}||\vec{n}_{1}|}|\psi_{1}\rangle\\
=\langle\psi_{2}|\hat{n}_{2}\cdot\hat{n}_{1}\mathbb{I}+i(\hat{n}_{2}\times\hat{n}_{1})\cdot\vec{\sigma}|\psi_{1}\rangle.
\end{multline}
Rearranging the terms on both sides of Eq.~\eqref{C3}, we obtain,
\begin{equation}\label{C4}
	(1-\hat{n}_{2}\cdot\hat{n}_{1})\langle\psi_{2}|\psi_{1}\rangle=i\langle\psi_{2}|(\hat{n}_{2}\times\hat{n}_{1})\cdot\vec{\sigma}|\psi_{1}\rangle.
\end{equation}
The RHS of Eq.~\eqref{C4} vanishes if the Bloch unit vectors $\hat{n}_{1}$ and $\hat{n}_{2}$ are parallel or anti-parallel to each other; if the LHS is to vanish  only when $\langle\psi_{2}|\psi_{1}\rangle=0$, then the condition for orthogonality of the Bloch vectors translates to,
\begin{equation}\label{C5}
	\hat{n}_{2}\cdot\hat{n}_{1}=-1,
\end{equation} 
implying the anti-parallel condition to be appropriate.

\section{Fidelity induced LO calculation for post slow quench unitary dynamics}\label{ap3}

The fidelity between two density matrices $\rho$ and $\sigma$ can be rewritten in the two dimensional basis of $|1\rangle_k$ and $|4\rangle_k$ as,
\begin{equation}\label{eq:B1}
 (F_1(\rho(0),\rho(t)))^2=\mathrm{Tr}\left[\rho(0)\rho(t)\right]+2\sqrt{(\det{\rho(0)})(\det{\rho(t)})}.
\end{equation}
 
 The first term on the RHS of Eq.~\eqref{eq:B1}, can be simplified as,
 \begin{multline}\label{B0}
 \mathrm{Tr}[\rho_k(0)\rho_k(t)]=\\
 \mathrm{Tr}\left[\sum_{i,j}p^k_ip^k_j|\psi_i^k(0)\rangle\langle\psi_i^k(0)|U_{k}(t)|\psi_j^k(0)\rangle\langle\psi_j^k(0)|U_{k}^{\dagger}(t)\right]\\
 =\sum_{i,j}p_i^kp^k_j|\langle\psi_{i}^k(0)|U_{k}(t)|\psi_j^k(0)\rangle|^2
 \end{multline}
 and the second term on the RHS of Eq.~\eqref{eq:B1} can be expanded in terms of the probabilities as,
 \begin{equation}\label{B}
 \sqrt{(\det{\rho_k(0)})(\det{\rho_k(t)})}= p_1^kp_4^k.
 \end{equation}
 Summing up Eq.~\eqref{B0} and Eq.~\eqref{B}, we obtain the fidelity in the reduced space spanned by the vectors $|1\rangle_k$ and $|4\rangle_k$.\\
 Now, as the space of $|2\rangle_k$ and $|3\rangle_k$ is decoupled from the space of $|1\rangle_k$ and $|4\rangle_k$ as explained in Appendix \ref{ap5}, the total fidelity separates as,
 \begin{equation}\label{eq:B2}
 F(\rho_k(0),\rho_k(t))=F_1(\rho_k(0),\rho_k(t))+F_2(\rho_k(0),\rho_k(t)),
 \end{equation}
 where $F_2$ is the fidelity between $\rho_k(0)$ and $\rho_k(t)$ in the basis of $|2\rangle_k$ and $|3\rangle_k$. Since, both the initial and the time evolved density matrices are diagonal in this basis and since, in the post quench unitary evolution the eigenvalues of the density matrix does not change with time,
 \begin{equation}\label{eq:B5}
 F_2=p_2^k+p_3^k.
 \end{equation}
Combining the results in Eq.~\eqref{eq:B2} and Eq.~\eqref{eq:B5}, we obtain the complete fidelity induced LO in Eq.~\eqref{eq:13} of the main text.\\

\section {Interferometric LO calculation for post slow quench unitary dynamics}\label{ap2}

The LO defined in Eq.~\eqref{eq:25} for the post quench unitary dynamics is 
\begin{equation}\label{eq:A1}
	\langle w_{k}(0)|w_{k}(t)\rangle=\sum_{i=1,4}p_{i}^k\langle \psi_{i}^{k}(0)|U_{k}(t)|\psi^{k}_{i}(0)\rangle +p_{2}^k+p^k_{3},
\end{equation} 
where $U_{k}(t)=\exp({-iH_{k}(h_f)t})$ and $H_{k}(h_f)$ is $2 \times 2$ final Hamiltonian in the  space spanned by basis spanned by $|1\rangle_k$ and $|4\rangle_k$. 
The first term on the RHS of Eq.~\eqref{eq:A1} can be re-written as $\mathrm{Tr}[\rho^{\prime}_{k}(0)U_{k}(t)]$ where $\rho^{\prime}_{k}$ is defined in Eq.~\eqref{eq:28} of the main text.
Using the identity $e^{iAx}=\mathbb{I}\cos{x}+iA\sin{x}$ where $x$ is a number and $A$ is a Hermitian matrix such that $A^{2}=\mathbb{I}$,
we  simplify  Eq.~\eqref{eq:A1} to the form:
\begin{equation}\label{eq:A2}
	\langle w_{k}(0)|w_{k}(t)\rangle=C_{k}\cos{\epsilon_{k}^{f}t}-i(\vec{n}_{k}\cdot\hat{d}_{f}(k))\sin{\epsilon_{k}^{f}t}+(1-C_{k}).
\end{equation}
Hence the critical $k^{*}$s can be determined from the condition,  
\begin{equation}\label{eq:A3}
	\vec{n}_{k^{*}}\cdot\hat{d}_{f}(k^{*})=0 
\end{equation}
and the corresponding critical times $t^{*}$s are obtained from,
\begin{equation}\label{eq:A4}
\begin{split}
C_{k^{*}}\cos{\epsilon_{k^{*}}^{f}t^{*}}+(1-C_{k^{*}})=0\\
\implies t^{*}=\frac{1}{\epsilon_{k^{*}}^{f}}\cos^{-1}{\left(1-\frac{1}{C_{k^{*}}}\right)}.
\end{split}
\end{equation}
The above critical times can only be obtained if $C_{k^{*}}\geq\frac{1}{2}$.


\begin{thebibliography}{99}




\bi{greiner02} M. Greiner , O. Mandel, T.  W. Hansch and  I. Bloch,  {\it Collapse and revival of the matter wave field of a BoseÐEinstein condensate},
Nature {\bf 419}, 51 (2002).    


\bi{kinoshita06} T. Kinoshita, T. Wenger and D. S. Weiss, {\it A quantum Newton's cradle},  Nature {\bf 440}, 900 (2006).    

\bi{fausti11} D. Fausti, R. I. Tobey, , N. Dean,  S. Kaiser, A. Dienst, M. C. Hoffmann, S. Pyon, T. Takayama, H. Takagi,4, A. Cavalleri,  {\it Light-Induced Superconductivity in a Stripe-Ordered Cuprate}, Science {\bf 331}, 189 (2011). 

\bi{gring12} M. Gring, M. Kuhnert, T. Langen, T. Kitagawa, B. Rauer, M. Schreitl, I. Mazets1, D. Adu Smith, E. Demler, and J. Schmiedmayer, {\it Relaxation and Pre-thermalization in an Isolated Quantum System}, 
Science {\bf 337}, 1318 (2012).   


\bi{trotzky12}  S. Trotzky,	Y-A. Chen,	A. Flesch,	I. P. McCulloch,	U. Schollwšck, J. Eisert and I. Bloch, {\it Probing the relaxation towards equilibrium in an isolated strongly correlated 1D Bose gas}, Nature {\bf 8}, 325 (2012).   





\bi{cheneau12}  M. Cheneau,	P. Barmettler,	D.  Poletti,	 M. Endres,	P.  Schauss,	T. Fukuhara,	C. Gross,	I. Bloch,	C.  Kollath	 and S.  Kuhr, {\it Light-cone-like spreading of correlations in a quantum many-body system}, Nature {\bf 481}, 484  (2012). 

\bi{rechtsman13}  M. C. Rechtsman,	J. M. Zeuner,	Y.  Plotnik,	 Y.  Lumer,	D.Podolsky,	F.  Dreisow,	S. Nolte,	M. Segev	and  A. Szameit, {\it Photonic Floquet topological insulators}, Nature
{\bf 496}  196 (2013).   



\bi{schreiber15}     M.  Schreiber,  S. S. Hodgman, P.  Bordia,  Henrik P. LŸschen, M. H. Fischer, R. Vosk, E. Altman, U. Schneider, I. Bloch, {\it 
Probing the relaxation towards equilibrium in an isolated strongly correlated 1D Bose gas}, Science {\bf 349}, 842 (2015).  





\bi{calabrese06} P. Calabrese, and  J. Cardy, {\it Time Dependence of Correlation Functions Following a Quantum Quench}, Phys. Rev. Lett. {\bf 96}, 136801 (2006).

\bi{rigol08} M. Rigol, V. Dunjko and M. Olshanii,  {\it Thermalization and its mechanism for generic isolated quantum systems}, Nature {\bf 452}, 854 (2008).   

\bibitem{oka09} T Oka, H Aoki, {\it Photovoltaic Hall effect in graphene}, Phys.  Rev.  B {\bf 79} 081406 (2009).   

\bibitem{mukherjee09} V. Mukherjee  and A. Dutta, {\it Effects of interference in the dynamics of a spin$- 1/2$ transverse XY chain driven periodically through quantum critical points}, J. Stat. Mech. P05005 (2009).  

\bi{bermudez09} A. Bermudez, D. Patane,  L. Amico, M. A. Martin-Delgado, {\it Topology-Induced Anomalous Defect Production by Crossing a Quantum Critical Point}, Phys. Rev. Lett. {\bf 102}, 135702, (2009).  

\bibitem{kitagawa10} T. Kitagawa, E. Berg, M. Rudner, and E. Demler, {\it Topological characterization of periodically driven quantum systems}, Phys. Rev. B
{\bf 82}, 235114 (2010).  

\bibitem{das10}
A. Das, {\it Exotic freezing of response in a quantum many-body system},  Phys. Rev. B {\bf 82}, 172402 (2010).  

\bibitem{pal10} A Pal and DA Huse, {\it Many-body localization phase transition}, Phys. Rev. B {\bf 82}, 174411  (2010).   

\bibitem{lindner11} N. H. Lindner, G. Refael and V. Galitski, {\it 
Floquet Topological Insulator in Semiconductor Quantum Wells}, Nat. Phys. {\bf 7}, 490-495, (2011).  

\bi{gambassi11} A. Gambassi and  A.  Silva, {\it 
Large deviations and universality in quantum quenches}, arXiv: 1106.2671 (2011); , Phys. Rev. Lett. {\bf 109}, 250602 (2012).


\bibitem{thakurathi13} M. Thakurathi, A. A. Patel, D. Sen, and A. Dutta, {\it Floquet topological phase transitions in a kicked Haldane-Chern insulator},  Phys. Rev. B {\bf 88}, 155133 (2013).  


\bibitem{Russomanno_PRL12} A. Russomanno,  A. Silva  and G. E. Santoro, {\it Multipartite entanglement after a quantum quench}, Phys. Rev.
Lett. {\bf 109}, 257201 (2012).   

\bi{sharma14}  S. Sharma, A. Russomanno, G. E. Santoro, and A. Dutta, {\it Loschmidt echo and dynamical fidelity in periodically driven quantum systems}, EPL {\bf 106},  67003 (2014).

\bi{nag14} T. Nag, S.  Roy, A. Dutta, and D.  Sen, {\it Dynamical localization in a chain of hard core bosons under periodic driving}, Phys. Rev. B {\bf 89}, 165425 (2014).

\bi{patel13} A. A. Patel, S. Sharma, A. Dutta, {\it Universal nonanalytic behavior of the Hall conductance in a Chern insulator at the
topologically driven nonequilibrium phase transition} ,Eur. Phys. Jour. B {\bf 86}, 367 (2013).

\bi{caio15} M. D. Caio, N. R. Cooper and M. J. Bhaseen, {\it Quantum Quenches in Chern Insulators}, Phys. Rev. Lett. {\bf 115}, 236403 (2015).    

\bi{nandkishore15} R. Nandkishore, D. A. Huse, {\it 
Many body localization and thermalization in quantum statistical mechanics}, Annual Review of Condensed Matter Physics,  {\bf 6}, 15-38 (2015).  


\bibitem{sen16} A.  Sen, S.  Nandy, K. Sengupta, {\it Entanglement generation in periodically driven integrable systems: Dynamical phase transitions and steady state}, Phys. Rev. B {\bf 94}, 214301 (2016).   

\bi{bukov16} 
M. Bukov, L. D'Alessio and A. Polkovnikov, {\it 
Universal High-Frequency Behavior of Periodically Driven Systems: from Dynamical Stabilization to Floquet Engineering}, Adv. Phys. {\bf 64} , No. 2, 139-226 (2016).   






\bibitem {dziarmaga10} J. Dziarmaga, {\it Dynamics of a Quantum Phase Transition and Relaxation to a Steady State}, Advances in Physics  {\bf 59}, 1063 (2010).  

\bibitem{polkovnikov11} A. Polkovnikov, K. Sengupta, A. Silva, and M. Vengalattore, {\it Colloquium: Nonequilibrium dynamics of closed interacting quantum systems}, Rev. Mod. Phys. {\bf 83}, 863 (2011).  



\bi{dutta15} A. Dutta, G. Aeppli, B. K. Chakrabarti, U. Divakaran, T. 
Rosenbaum and D. Sen, \textit{Quantum Phase Transitions in Transverse Field 
	Spin Models: From Statistical Physics to Quantum Information} (Cambridge 
University Press, Cambridge, 2015).    

\bi{eisert15} J. Eisert, M. Friesdorf and C. Gogolin, {\it 
Quantum many-body systems out of equilibrium}, Nat. Phys. {\bf 11}, 124 (2015).  

\bi{alessio16} L. D'Alessio, Y.  Kafri, A. Polkovnikov, M. Rigol, {\it 
From Quantum Chaos and Eigenstate Thermalization to Statistical Mechanics and Thermodynamics}, Adv. Phys. {\bf 65}, 239 (2016).  

\bi{jstat} J.  Stat.  Mech.: Theo. and Expt,   special issue{\it Quantum Integrability in Out of Equilibrium Systems} edited by  P. Calabrese., F. H. L. Essler and G. Mussardo, {\bf 2016} (2016).
 
\bi{heyl13} M. Heyl, A. Polkovnikov, and S. Kehrein, {\it Dynamical Quantum Phase Transitions in the Transverse-Field Ising Model}, Phys. Rev. Lett., {\bf 110}, 135704 (2013).   


\bi {fisher65} M.E. Fisher, in {\it Boulder Lectures in Theoretical Physics} (University of Colorado, Boulder, 1965), Vol. 7. 

\bi {lee52} C. Yang and T. Lee, {\it Statistical Theory of Equations of State and Phase Transitions. I. Theory of Condensation}, Phys. Rev. {\bf 87}, 404 (1952). 



\bi{saarloos84} W. van Saarloos and D. Kurtze, {\it  Location of Zeros in the Complex Temperature Plane: Absence of Lee-Yang Theorem}, J. Phys. A {\bf 17}, 1301 (1984).

\bi{jurcevic16} P. Jurcevic, H. Shen, P. Hauke, C. Maier, T. Brydges, C. Hempel, B. P. Lanyon, M. Heyl, R. Blatt, C. F. Roos, {\it Direct Observation of Dynamical Quantum Phase Transitions in an Interacting Many-Body System}, Phys. Rev. Lett. {\bf119}, 080501 (2017).

\bi{flaschner} Flaschner, D. Vogel, M. Tarnowski, B, S. Rem, D.-S. Luhmann, M. Heyl, J. Budich, L. Mathey, K. Sengstock,
C. Weitenberg, {\it Experimental reconstruction of the Berry curvature in a Floquet Bloch band}, Nat. Phys. 1745-2481 (2017). 





  





\bi{karrasch13} C. Karrasch and D. Schuricht, {\it Dynamical phase transitions after quenches in nonintegrable models}, Phys. Rev. B, {\bf 87}, 195104 (2013).  

\bi{kriel14} N. Kriel, C. Karrasch, and S. Kehrein, {\it Dynamical quantum phase transitions in the axial next-nearest-neighbor Ising chain}, Phys. Rev. B {\bf 90}, 125106 (2014).

\bi{andraschko14} F. Andraschko, J. Sirker, {\it Dynamical quantum phase transitions and the Loschmidt echo: A transfer matrix approach}, Phys. Rev. B {\bf 89}, 125120 (2014).  

\bi{canovi14} E. Canovi, P. Werner, and M. Eckstein, {\it First-Order Dynamical Phase Transitions}, Phys. Rev. Lett. {\bf 113}, 265702 (2014).

\bi{heyl14} M. Heyl, {\it Dynamical Quantum Phase Transitions in Systems with Broken-Symmetry Phases}, Phys. Rev. Lett., {\bf 113}, 205701 (2014).

\bi{vajna14} S. Vajna and B. Dora, {\it Disentangling dynamical phase transitions from equilibrium phase transitions}, Phys. Rev. B {\bf 89}, 161105(R) (2014).

\bi{vajna15} S. Vajna and B. Dora, {\it Topological classification of dynamical phase transitions}, Phys. Rev. B {\bf 91}, 155127 (2015).

\bi{heyl15} M. Heyl, {\it Scaling and Universality at Dynamical Quantum Phase Transitions}, Phys. Rev. Lett., {\bf 115}, 140602 (2015) .

\bi{budich15} J. C. Budich and  M. Heyl, {\it Dynamical topological order parameters far from equilibrium}, Phys. Rev. B {\bf 93}, 085416 (2016).     



\bi{palami15} T. Palmai, {\it Edge exponents in work statistics out of equilibrium and dynamical phase transitions from scattering theory in one-dimensional gapped systems}, Phys. Rev. B {\bf 92}, 235433 (2015).  

\bi{schmitt15} M. Schmitt and S. Kehrein, {\it Dynamical Quantum Phase Transitions in the Kitaev Honeycomb Model}, Phys. Rev. B {\bf 92}, 075114 (2015).

\bi{divakaran16} U. Divakaran, S. Sharma and A. Dutta, {\it Tuning the presence of dynamical phase transitions in a generalized XY spin chain}, Phys. Rev. E {\bf 93}, 052133 (2016). 


\bi{huang16} Z.  Huang, and A.  V. Balatsky, {\it Dynamical Quantum Phase Transitions: Role of Topological Nodes in Wave Function Overlaps}, Phys. Rev. Lett. {\bf 117}, 086802 (2016).   

\bi{puskarov16} T. Puskarov and D. Schuricht, {\it Time evolution during and after finite-time
quantum quenches in the transverse-field Ising chain}, arXiv: 1608.05584 (2016).  

\bi{zhang16} J. M. Zhang abd  H.-T. Yang, {\it Sudden jumps and plateaus in the quench dynamics of a Bloch state}, arXiv: 1605.05403 (2016). 

\bi{heyl16} M. Heyl, {\it Quenching a quantum critical state by the order parameter: Dynamical quantum phase transitions and quantum speed limits}, Phys. Rev. B {\bf 95}, 060504 (2017). 

\bi{zunkovic16} B. Zunkovic, A. Silva, M. Fabrizio, {\it Dynamical phase transitions and Loschmidt echo in the infinite-range XY model}, Philos Trans A Math Phys Eng Sci. 374(2069), (2016).




\bi{sei17} T.  Obuchi, S. Suzuki, K. Takahashi, {\it Complex semiclassical analysis of the Loschmidt amplitude and
dynamical quantum phase transitions}, arXiv:1702.05396 (2017). 

\bi{bhattacharya1} Utso Bhattacharya and Amit Dutta, {\it Emergent topology and dynamical quantum phase transitions in two-dimensional closed quantum systems}, Phys. Rev. B {\bf96}, 014302 (2017). 


\bi{bhattacharya2} Utso Bhattacharya and Amit Dutta, {\it Interconnections between equilibrium topology and dynamical quantum phase transitions in a linearly ramped Haldane model}, Phys. Rev. B 95, 184307 (2017).


\bi{fogarty17} Thom‡s Fogarty, Ayaka Usui, Thomas Busch, Alessandro Silva, John Goold, {\it 
Dynamical phase transitions, temporal orthogonality and the dynamics of observables in one dimensional ultra-cold quantum gases: from the continuum to the lattice}, New J. Phys. 19 113018 (2017). 

\bi{halimeh17} J. C. Halimeh and V. Zauner-Stauber, {\it Dynamical phase diagram of quantum spin chains with long-range interactions}, Phys. Rev. B {\bf96}, 134427 (2017).

\bi{homri17} I. Homrighausen, N. O. Abeling, V. Zauner-Stauber, and J. C.Halimeh, , {\it Anomalous dynamical phase in quantum spin chains with long-range interactions}, Phys.Rev.B {\bf 96}, 104436 (2017).


\bi{dutta17} A. Dutta and A. Dutta, {\it Probing the role of long-range interactions in the dynamics of a long-range Kitaev chain}, Phys. Rev. B {\bf 96}, 125113 (2017).


\bi{mera17}Bruno Mera, Chrysoula Vlachou, Nikola Paunkovic, V'tor R. Vieira, Oscar Viyuela, {\it Dynamical phase transitions at finite temperature from fidelity and interferometric Loschmidt echo induced metrics}, arXiv:1712.01314 (2017)

\bi{sedlmayr181}N. Sedlmayr, P. JŠger, M. Maiti, J. Sirker, {\it A bulk-boundary correspondence for dynamical phase transitions in one-dimensional topological insulators and superconductors}, Phys. Rev. B {\bf97}, 064304 (2018) 



\bi{heyl18} Daniele Trapin, Markus Heyl, {\it 
Constructing effective free energies for dynamical quantum phase transitions in the transverse-field Ising chain}, arXiv:1802.00020 (2018).

\bi{bhattacharjee18} S. Bhattacharjee and A. Dutta, {\it Dynamical Quantum Phase Transitions in Extended Transverse Ising Models}, arXiv:1801.09463 (2018).

\bi{kennes18} D. M. Kennes, D. Schuricht, C. Karrasch, {\it Controlling Dynamical Quantum Phase Transitions},  arXiv:1803.09242 (2018).




\bi{pollmann10} F. Pollmann, S. Mukerjee, A. G. Green, and J. E. Moore, {\it Dynamics after a sweep through a quantum critical point}, Phys. Rev. E {\bf 81}, 020101(R) (2010).

\bi{sharma15} S. Sharma, S. Suzuki and A. Dutta, {\it Quenches and dynamical phase transitions in a nonintegrable quantum Ising model}, Phys. Rev. B {\bf 92}, 104306 (2015). 



\bi{sharma16} S. Sharma, U. Divakaran, A. Polkovnikov and A. Dutta, {\it 
Slow quenches in a quantum Ising chain; dynamical phase transitions and topology}, Phys. Rev. B {\bf 93}, 144306 (2016).  






\bi{bandyopadhyay17} U. Bhattacharya, S. Bandyopadhyay and A. Dutta, {\it Mixed state dynamical quantum phase transitions}, Phys. Rev. B {\bf96}, 180303(R) (2017).    

\bi{heyl_mixed17} M.Heyl and J.C. Budich, {\it Dynamical topological quantum phase transitions for mixed states}, Phys. Rev. B {\bf96}, 180304(R) (2017) . 

\bi{abeling16} Nils O. Abeling, Stefan Kehrein, {\it Quantum quench dynamics in the transverse field Ising model at nonzero temperatures}, Phys. Rev. B {\bf 93}, 104302 (2016).

\bi{sedlmayr182} N. Sedlmayr, M. Fleischhauer, J. Sirker, {\it The fate of dynamical phase transitions at finite temperatures and in open systems}, Phys. Rev. B {\bf97}, 045147 (2018)(2018).



\bi{heyl17} Markus Heyl, {\it Dynamical quantum phase transitions: a review}, arXiv:1709.07461 (2017).

\bi{victor17} Victor Gurarie, {\it Quantum Phase Transitions Go
Dynamical}10.1103/Physics.10.95 (2017).

\bi{zvyagin17} A.A. Zvyagin, {\it 
Dynamical quantum phase transitions}, Low Temp. Phys. {\bf 42}, 971 (2016).



\bi{dorner12} R. Dorner,J. Goold, C. Cormick, M. Paternostro,and V. Vedral, {\it Emergent Thermodynamics in a Quenched Quantum Many-Body System}, Phys. Rev. Lett. {\bf 109}, 160601 (2012).

\bi{sharma_entropy15} Shraddha Sharma and Amit Dutta, {\it One and two-dimensional quantum models: quenches and the scaling of irreversible entropy}, Phys. Rev. E {\bf 92}, 022108 (2015).

\bi{fazio17}  Maximilian Keck , Simone Montangero , Giuseppe E Santoro , Rosario Fazio  and Davide Rossini, {\it 
Dissipation in adiabatic quantum computers: Lessons from an exactly solvable model}, New Journal of Physics {\bf 19}, (2017). 

\bibitem{sachdev10} S. Sachdev, {\it  Quantum Phase Transitions} (Cambridge University Press, Cambridge, UK, 2010).  
%
\bibitem{suzuki13} S. Suzuki, J-i Inoue and Bikas K. Chakrabarti, {\it Quantum Ising Phases and Transitions in Transverse Ising Models}  (Springer, Lecture Notes in Physics, Vol. 862 (2013)).         
%

\bi{mixed_fidelity} P. Zanardi, H. T. Quan, Xiaoguang Wang, and C. P. Sun, {\it Mixed-state fidelity and quantum criticality at finite temperature}, Phys. Rev. A 75, 032109 (2007).

\bi{uhlmann86} A. Uhlmann, Rep. Math. Phys. {\bf 24}, 229 (1986).

\bi{uhlmann89}  A. Uhlmann, Ann. Phys. 501, 63 (1989).

\bi{sjoqvist} E. Sjšqvist, A. K. Pati, A. Ekert, J. S. Anandan, M. Ericsson,
D. K. L. Oi, and V. Vedral, {\it Geometric Phases for Mixed States in Interferometry}, Phys. Rev. Lett. 85, 2845 (2000).


\bi{breuer_book} H.-P. Breuer and  F.  Petruccione, {\it Theory of open quantum systems}, Oxford University Press, Oxford (2002).

\bi{thekkadath16} G. S. Thekkadath, L. Giner, Y. Chalich, M. J. Horton, J. Banker, and J. S. Lundeen, {\it Direct measurement of the density matrix of a quantum system}, Phys. Rev. Lett. 117, 120401 (2016).







\end{thebibliography}
\end{document}